\documentclass[aps,preprint]{revtex4}%
\usepackage{amsfonts}
\usepackage{amsmath}
\usepackage{amssymb}
\usepackage{graphicx}%
\begin{document}
\title{The symmetric heavy-light ansatz}
\author{Dean Lee}
\affiliation{Department of Physics, North Carolina State University, Raleigh, NC 27695}

\begin{abstract}
The symmetric heavy-light ansatz is a method for finding the ground state of
any dilute unpolarized system of attractive two-component fermions.
\ Operationally it can be viewed as a generalization of the Kohn-Sham
equations in density functional\ theory applied to $N$-body density
correlations. \ While the original Hamiltonian has an exact $Z_{2}$ symmetry,
the heavy-light ansatz breaks this symmetry by skewing the mass ratio of the
two components. \ In the limit where one component is infinitely heavy, the
many-body problem can be solved in terms of single-particle orbitals. \ The
original $Z_{2}$ symmetry is recovered by enforcing $Z_{2}$ symmetry as a
constraint on $N$-body density correlations for the two components.\ \ For the
1D, 2D, and 3D attractive Hubbard models the method is in very good agreement
with exact Lanczos calculations for few-body systems at arbitrary coupling.
\ For the 3D\ attractive Hubbard model there is very good agreement with
lattice Monte Carlo results for many-body systems in the limit of infinite
scattering length.

\end{abstract}
\maketitle

\section{Introduction}

Dilute two-component fermions with attractive interactions have some universal
features of relevance to several areas of physics. \ In three dimensions the
unpolarized ground state is believed to be superfluid with properties in
between a Bardeen-Cooper-Schrieffer (BCS) fermionic superfluid at weak
coupling and a Bose-Einstein condensate (BEC) of bound dimers at strong
coupling \cite{Eagles:1969PR,Leggett:1980pro,Nozieres:1985JLTP}. \ The
crossover transition occurs somewhere near the unitarity point, a
scale-invariant point where the range of the interaction is zero and the
scattering length is infinite. \ Much recent interest on this topic has been
stimulated by experimental progress with cold atomic Fermi gases. \ Starting
with a dilute Fermi gas, where the effective range of the interaction is
negligible compared with the interparticle spacing, the scattering length can
be tuned using a magnetic Feshbach resonance
\cite{O'Hara:2002,Gupta:2002,Regal:2003,Bourdel:2003,Gehm:2003,Bartenstein:2004,Kinast:2005}%
. \ In nuclear physics the phenomenology of the unitarity point is relevant to
cold dilute neutron matter. \ The neutron scattering length is roughly $-18$
fm while the range of the interaction is comparable to the Compton wavelength
of the pion, $m_{\pi}^{-1}\approx1.4$ fm. \ Therefore the unitarity point is
approximately realized when the interparticle spacing is about $5$ fm. \ This
range of neutron density is expected in the inner crust of neutron stars.

In this paper we introduce a new theoretical method called the symmetric
heavy-light ansatz. \ The ansatz is used to find the ground state energy of
dilute unpolarized two-component fermions with attractive interactions.
\ While our main interest is the three-dimensional system, we demonstrate that
the ansatz is accurate for any number of spatial dimensions. \ Operationally
it can be described as an $N$-body generalization of the Kohn-Sham equations
in density functional\ theory. \ But rather than solving for one-body
densities, we use the symmetric heavy-light ansatz to determine $N$-body
density correlations. \ While useful as a numerical technique, the ansatz also
provides a heuristic picture of the underlying competition between Fermi
repulsion and attractive interactions.

We start with an effective Hamiltonian describing two-component fermions with
short-range attractive interactions. \ For the unpolarized case the ground
state has an exact $Z_{2}$ symmetry corresponding with interchanging
components. \ The first step of the ansatz is to break this $Z_{2}$ symmetry
by skewing the mass ratio of the two components. \ In fact we take the extreme
limit where one component is infinitely heavy. \ In this limit the many-body
problem is completely solved in terms of single-particle orbitals. \ The
original $Z_{2}$ symmetry is reintroduced as a constraint on the $N$-body
correlations for the two components.\ \ In the simple approximation where only
the lowest orbitals are filled, this constraint completely determines the
$N$-body density correlations for each fermion component. \ We run the method
through several numerical tests and find very good agreement for a number of
exact Lanczos and Monte Carlo results for the attractive Hubbard model in one,
two, and three dimensions. \ The low computational scaling of the ansatz
allows us to make predictions for much larger systems which are not yet
accessible using alternate methods.

The organization of the paper is as follows. \ The beginning is a short
summary of symmetries of the effective Hamiltonian. \ After this we introduce
the heavy-light Hamiltonian $H_{\text{HL}}$ and the symmetric heavy-light
ansatz. \ Some results relating to heavy-light orbitals are derived, and the
spectrum of the heavy-light Hamiltonian $H_{\text{HL}}$ is explored
numerically for the unpolarized four-body system in three dimensions. \ This
analysis provides motivation for the lowest filling approximation and the
construction of $N$-body fixed-point densities. \ We then design a Markov
chain algorithm to generate fixed-point densities. \ As a first test we
compare the results of the symmetric heavy-light ansatz with exact Lanczos
results for the four-body system in three dimensions. \ We also compare with
exact results for four- and six-body systems in both one and two dimensions.
\ We then consider larger systems in three dimensions at unitarity and the
ground state energy for general values of the scattering length. \ We conclude
with a discussion of some extensions of the method and future work.

\section{Symmetries of the Hamiltonian}

In continuum notation we can write the effective Hamiltonian for two-component
fermions in $d\leq3$ dimensions with short-range interactions as%
\begin{equation}
H=-\frac{1}{2m}\sum_{\sigma=\uparrow,\downarrow}\int d^{d}\vec{r}\;a_{\sigma
}^{\dagger}(\vec{r})\vec{\nabla}^{2}a_{\sigma}(\vec{r})+C\int d^{d}\vec
{r}\;a_{\downarrow}^{\dagger}(\vec{r})a_{\uparrow}^{\dagger}(\vec
{r})a_{\uparrow}(\vec{r})a_{\downarrow}(\vec{r}). \label{continuum}%
\end{equation}
Our main interest is the case $d=3$, but we also consider lower dimensional
systems $d=1,2$. $\ a_{\sigma}$ and $a_{\sigma}^{\dagger}$ are annihilation
and creation operators for fermions with two components. \ We refer to these
components as up and down spins.\ The mass of the fermion is $m$, and the
coefficient $C$ is assumed to be negative so that the interaction is
attractive. \ The strength of $C$ depends on the scheme used to regulate the
short distance behavior. \ This Hamiltonian has a global $U(1)$ fermion-number
symmetry,%
\begin{equation}
\left[
\genfrac{}{}{0pt}{}{a_{\uparrow}(\vec{r})}{a_{\downarrow}(\vec{r})}%
\right]  \rightarrow e^{i\phi}\left[
\genfrac{}{}{0pt}{}{a_{\uparrow}(\vec{r})}{a_{\downarrow}(\vec{r})}%
\right]  ,
\end{equation}
where $\phi$ is any real constant. \ It also has a global $SU(2)$ spin
symmetry%
\begin{equation}
\left[
\genfrac{}{}{0pt}{}{a_{\uparrow}(\vec{r})}{a_{\downarrow}(\vec{r})}%
\right]  \rightarrow e^{i\vec{\phi}\cdot\vec{\sigma}}\left[
\genfrac{}{}{0pt}{}{a_{\uparrow}(\vec{r})}{a_{\downarrow}(\vec{r})}%
\right]  ,
\end{equation}
where $\vec{\sigma}$ denotes the $2\times2$ Pauli spin matrices and $\vec
{\phi}$ is any constant real three-component vector. \ Since there is no
coupling between spin and orbital angular momentum, this $SU(2)$ symmetry
should be regarded as an internal symmetry decoupled from spatial rotations.

The lowest-dimensional local bosonic operator that can be constructed from the
annihilation field operators is%
\begin{equation}
\psi^{2}(\vec{r})=a_{\uparrow}(\vec{r})a_{\downarrow}(\vec{r}).
\end{equation}
We note that $\psi^{2}$ is invariant under the $SU(2)$ spin symmetry but phase
rotates under the $U(1)$ fermion-number symmetry,
\begin{equation}
\psi^{2}(\vec{r})\rightarrow e^{2i\phi}\psi^{2}(\vec{r}).
\end{equation}
Therefore if there is some critical temperature below which $\psi^{2}$ has
long-range spatial correlations,
\begin{equation}
\lim_{\left\vert \vec{r}\right\vert \rightarrow\infty}\left\langle
\psi^{2\dagger}(\vec{r})\psi^{2}(\vec{0})\right\rangle \neq0,
\end{equation}
then the $U(1)$ fermion-number symmetry is spontaneously broken. \ This
condition of off-diagonal long-range order \cite{Penrose:1956,Gorkov:1958} is
the standard definition for superfluidity with S-wave pairing. \ We observe
that the $SU(2)$ spin symmetry is not broken by the expectation value of
$\psi^{2}$.

\section{Symmetric heavy-light ansatz}

Let $K_{\uparrow}$ and $K_{\downarrow}$ be the kinetic energy operators
associated with the up and down spins respectively,%
\begin{align}
K_{\uparrow}  &  =-\frac{1}{2m}\int d^{d}\vec{r}\;a_{\uparrow}^{\dagger}%
(\vec{r})\vec{\nabla}^{2}a_{\uparrow}(\vec{r}),\\
K_{\downarrow}  &  =-\frac{1}{2m}\int d^{d}\vec{r}\;a_{\downarrow}^{\dagger
}(\vec{r})\vec{\nabla}^{2}a_{\downarrow}(\vec{r}).
\end{align}
We define $H_{\text{HL}}$ as%
\begin{equation}
H_{\text{HL}}=H-K_{\uparrow}+K_{\downarrow}=2K_{\downarrow}+V,
\end{equation}
and refer to $H_{\text{HL}}$ as the heavy-light Hamiltonian. \ In
$H_{\text{HL}}$ we have deleted the kinetic energy for the up spin while
doubling the kinetic energy of the down spin. \ This could be viewed as
introducing spin-dependent masses $m_{\uparrow}$ and $m_{\downarrow}$. \ For
the original Hamiltonian $H$ we have $m_{\uparrow}=m_{\downarrow}=m$, while in
$H_{\text{HL}}$ we have $m_{\uparrow}=\infty,m_{\downarrow}=m/2.$

The physics of the two-body system in the center-of-mass frame is exactly the
same for $H$ and $H_{\text{HL}}$. \ The reduced mass $\mu$ defined by%
\begin{equation}
\frac{1}{\mu}=\frac{1}{m_{\uparrow}}+\frac{1}{m_{\downarrow}}%
\end{equation}
equals $m/2$ in both cases. \ The exact equivalence of $H$ and $H_{\text{HL}}$
for the center-of-mass two-body system is preserved by most regularization
schemes such as dimensional regularization, momentum cutoff schemes, and
Hamiltonian lattice regularization. \ Hamiltonian lattice regularization is
discussed in the appendix.

While $H_{\text{HL}}$ and $H$ are identical for the two-body system, they are
very different for more than two particles. $\ $For example when the
ultraviolet cutoff scale goes to infinity for the three-dimensional system,
$H$ has a well-defined continuum limit while $H_{\text{HL}}$ is unbounded
below due to a clustering instability. \ The simplest example where this
occurs is the three-body system consisting of two up spins and one down spin.
\ It is known that this system collapses due to an attractive $1/r^{2}$
potential when $m_{\uparrow}\gtrsim13\,m_{\downarrow}$ \cite{Petrov:2002a}.
\ On the other hand the clustering instability in $H_{\text{HL}}$ is
eliminated when we project onto quantum states invariant under the interchange
of up and down spins. \ In the following we show that the minimum expectation
value for $H_{\text{HL}}$ restricted to this up-down symmetric space equals
the ground state energy of $H$.

We denote a state with $N_{\uparrow}$ up spins and $N_{\downarrow}$ down spins
as an $N_{\uparrow},N_{\downarrow}$ state. \ We also specify the total
momentum $\vec{P}$ and total spin $S$ of the $SU(2)$ spin representation.
\ Let $\left\vert \Psi_{N,N}^{0}\right\rangle $ be the normalized ground state
of $H$ for the $N,N$ system in a periodic cube of length $L$. \ Since the
$SU(2)$ spin symmetry is not broken by the expectation value of $\psi^{2}$, we
assume that $\left\vert \Psi_{N,N}^{0}\right\rangle $ lies in the
spin-invariant sector $S=0$. \ Let $E_{N,N}^{0}$ be the corresponding ground
state energy. \ Let $U$ be any unitary operator which interchanges the up and
down spins through a $\pi$-radian spin rotation. \ Without loss of generality
we take $U$ to be%
\begin{equation}
U=\exp\left[  -i\frac{\pi}{2}\int d^{d}\vec{r}\;a_{\downarrow}^{\dagger}%
(\vec{r})a_{\uparrow}(\vec{r})-i\frac{\pi}{2}\int d^{d}\vec{r}\;a_{\uparrow
}^{\dagger}(\vec{r})a_{\downarrow}(\vec{r})\right]  .
\end{equation}
Clearly%
\begin{equation}
U^{\dagger}a_{\uparrow}U=-ia_{\downarrow}, \label{Uup}%
\end{equation}%
\begin{equation}
U^{\dagger}a_{\downarrow}U=-ia_{\uparrow}. \label{Udown}%
\end{equation}
The phases appearing in Eq.~(\ref{Uup}) and (\ref{Udown}) are necessary in
order that the pair annihilation operator $\psi^{2}(\vec{r})$ remains
invariant,%
\begin{equation}
U^{\dagger}\psi^{2}(\vec{r})U=U^{\dagger}a_{\uparrow}(\vec{r})a_{\downarrow
}(\vec{r})U=-a_{\downarrow}(\vec{r})a_{\uparrow}(\vec{r})=\psi^{2}(\vec{r}).
\end{equation}
Any state with an even number of fermions is invariant under two successive
transformations of $U$. \ Therefore $U$ acting on the space of even fermion
states generates a unitary representation of the group $Z_{2}$.

Let $\left\vert \Phi_{N,N}\right\rangle $ be any $N,N$ state which is $Z_{2}$
invariant,%
\begin{equation}
U\left\vert \Phi_{N,N}\right\rangle =\left\vert \Phi_{N,N}\right\rangle
\text{.}%
\end{equation}
Since%
\begin{equation}
\left\langle \Phi_{N,N}\right\vert K_{\uparrow}\left\vert \Phi_{N,N}%
\right\rangle =\left\langle \Phi_{N,N}\right\vert K_{\downarrow}\left\vert
\Phi_{N,N}\right\rangle ,
\end{equation}
it follows that%
\begin{equation}
\left\langle \Phi_{N,N}\right\vert H_{\text{HL}}\left\vert \Phi_{N,N}%
\right\rangle =\left\langle \Phi_{N,N}\right\vert H\left\vert \Phi
_{N,N}\right\rangle .
\end{equation}
In particular this means%
\begin{equation}
\min_{U\left\vert \Phi_{N,N}\right\rangle =\left\vert \Phi_{N,N}\right\rangle
}\frac{\left\langle \Phi_{N,N}\right\vert H_{\text{HL}}\left\vert \Phi
_{N,N}\right\rangle }{\left\langle \Phi_{N,N}\right.  \left\vert \Phi
_{N,N}\right\rangle }=\min_{U\left\vert \Phi_{N,N}\right\rangle =\left\vert
\Phi_{N,N}\right\rangle }\frac{\left\langle \Phi_{N,N}\right\vert H\left\vert
\Phi_{N,N}\right\rangle }{\left\langle \Phi_{N,N}\right.  \left\vert
\Phi_{N,N}\right\rangle }.
\end{equation}
Since $\left\vert \Psi_{N,N}^{0}\right\rangle $ is a $Z_{2}$-invariant state
we conclude that%
\begin{equation}
\min_{U\left\vert \Phi_{N,N}\right\rangle =\left\vert \Phi_{N,N}\right\rangle
}\frac{\left\langle \Phi_{N,N}\right\vert H_{\text{HL}}\left\vert \Phi
_{N,N}\right\rangle }{\left\langle \Phi_{N,N}\right.  \left\vert \Phi
_{N,N}\right\rangle }=E_{N,N}^{0}. \label{heavylightansatz}%
\end{equation}
We refer to this exact relation as the symmetric heavy-light ansatz. \ The
task of computing the ground state energy $E_{N,N}^{0}$ is reduced to finding
the minimum of value of the Rayleigh-Ritz ratio%
\begin{equation}
\frac{\left\langle \Phi_{N,N}\right\vert H_{\text{HL}}\left\vert \Phi
_{N,N}\right\rangle }{\left\langle \Phi_{N,N}\right.  \left\vert \Phi
_{N,N}\right\rangle }%
\end{equation}
under the constraint of $Z_{2}$ invariance,%
\begin{equation}
U\left\vert \Phi_{N,N}\right\rangle =\left\vert \Phi_{N,N}\right\rangle .
\end{equation}

\section{Heavy-light orbitals}

In this section we discuss some properties of the energy eigenstates of the
heavy-light Hamiltonian $H_{\text{HL}}$. \ Since the up spins are infinitely
massive, we can fix these particles at locations $\vec{R}_{1},\vec{R}%
_{2},\cdots,\vec{R}_{N}$. \ Due to antisymmetry each $\vec{R}_{i}$ must be
distinct, and we use the shorthand notation%
\begin{equation}
\mathbf{R}=\left\{  \vec{R}_{1},\vec{R}_{2},\cdots,\vec{R}_{N}\right\}
\end{equation}
for the unordered set of vectors. \ We define $\left\vert \mathbf{R}%
\right\rangle $ as the corresponding antisymmetric state of localized
particles,%
\begin{equation}
\left\vert \mathbf{R}\right\rangle =\frac{1}{\sqrt{N!}}\sum_{\pi}%
sgn(\pi)\left\vert \vec{R}_{\pi(1)}\right\rangle \otimes\left\vert \vec
{R}_{\pi(2)}\right\rangle \otimes\cdots\left\vert \vec{R}_{\pi(N)}%
\right\rangle .
\end{equation}
The summation is over all permutations $\pi$ of the integers $1,2,\cdots,N$.
$\ sgn(\pi)$ equals $-1$ for odd permutations and $+1$ for even permutations.
\ The normalization for $\left\vert \mathbf{R}\right\rangle $ is then simply%
\begin{equation}
\left\langle \mathbf{R}^{\prime}\right.  \left\vert \mathbf{R}\right\rangle
=\det\Delta_{ij}(\mathbf{R}^{\prime}\mathbf{,R),}%
\end{equation}
where%
\begin{equation}
\Delta_{ij}(\mathbf{R}^{\prime}\mathbf{,R)}=\delta^{(d)}\left(  \vec{R}%
_{i}^{\prime}-\vec{R}_{j}\right)  . \label{D_normalization}%
\end{equation}
Later in the discussion when we consider lattice models the Dirac delta
functions will be replaced by Kronecker delta functions.

For any given $\mathbf{R}$ the down spins see a static delta function
potential from each up spin,%
\begin{equation}
\sum_{i=1,\cdots,N}C\delta^{(d)}(\vec{r}-\vec{R}_{i})\text{.}%
\end{equation}
For the very special case where $\mathbf{R}$ is a regular cubic array, this
system is known as the Kronig-Penney model \cite{Kronig:1931}. \ In general
though $\mathbf{R}$ is irregular without translational and rotational
symmetries. \ For a given $\mathbf{R}$ let the normalized single-particle
orbitals be $\left\vert f_{j}(\mathbf{R})\right\rangle $ with corresponding
eigenvalues $E_{j}(\mathbf{R})$. \ By convention we label the orbitals so that
$E_{j}(\mathbf{R})$ increases with $j,$%
\begin{equation}
E_{1}(\mathbf{R})\leq E_{2}(\mathbf{R})\leq\cdots\leq E_{j}(\mathbf{R}%
)\leq\cdots\text{.}%
\end{equation}
For a single orbital we write the position-space wavefunction as%
\begin{equation}
f_{j}(\vec{r},\mathbf{R})=\left\langle \vec{r}\right.  \left\vert
f_{j}(\mathbf{R})\right\rangle .
\end{equation}
Analogous to $\mathbf{R}$, we define%
\begin{equation}
\mathbf{r}=\left\{  \vec{r}_{1},\vec{r}_{2},\cdots,\vec{r}_{N}\right\}
\end{equation}
for the unordered set of distinct vectors $\vec{r}_{i}$. \ These correspond
with possible locations for the $N$ down spins. \ We also\ define the
antisymmetric product of position eigenstates%
\begin{equation}
\left\vert \mathbf{r}\right\rangle =\frac{1}{\sqrt{N!}}\sum_{\pi}%
sgn(\pi)\left\vert \vec{r}_{\pi(1)}\right\rangle \otimes\left\vert \vec
{r}_{\pi(2)}\right\rangle \otimes\cdots\left\vert \vec{r}_{\pi(N)}%
\right\rangle .
\end{equation}
with normalization%
\begin{equation}
\left\langle \mathbf{r}^{\prime}\right.  \left\vert \mathbf{r}\right\rangle
=\det\Delta_{ij}(\mathbf{r}^{\prime}\mathbf{,r).}%
\end{equation}

Since there is no interaction between down spins, each eigenstate of
$H_{\text{HL}}$ for fixed $\mathbf{R}$ is an antisymmetric product of
single-particle orbitals. \ Let us write the normalized quantum state with $N$
down spins filling orbitals $j_{1},\cdots,j_{N}$ as%
\begin{equation}
\bigwedge\limits_{n=1,\cdots,N}\left\vert f_{j_{n}}(\mathbf{R})\right\rangle
=\frac{1}{\sqrt{N!}}\sum_{\pi}sgn(\pi)\left\vert f_{\pi(j_{1})}(\mathbf{R}%
)\right\rangle \otimes\left\vert f_{\pi(j_{2})}(\mathbf{R})\right\rangle
\otimes\cdots\left\vert f_{\pi(j_{N})}(\mathbf{R})\right\rangle .
\end{equation}
The position state wavefunction for this state is a Slater determinant,%
\begin{equation}
\left\langle \mathbf{r}\right\vert \bigwedge\limits_{n=1,\cdots,N}\left\vert
f_{j_{n}}(\mathbf{R})\right\rangle =\det\left[
\begin{array}
[c]{cccc}%
f_{j_{1}}(\vec{r}_{1},\mathbf{R}) & f_{j_{2}}(\vec{r}_{1},\mathbf{R}) & \cdots
& f_{j_{N}}(\vec{r}_{1},\mathbf{R})\\
f_{j_{1}}(\vec{r}_{2},\mathbf{R}) & f_{j_{2}}(\vec{r}_{2},\mathbf{R}) & \cdots
& f_{j_{N}}(\vec{r}_{2},\mathbf{R})\\
\vdots & \vdots & \ddots & \vdots\\
f_{j_{1}}(\vec{r}_{N},\mathbf{R}) & f_{j_{2}}(\vec{r}_{N},\mathbf{R}) & \cdots
& f_{j_{N}}(\vec{r}_{1},\mathbf{R})
\end{array}
\right]  .
\end{equation}
All eigenstates of $H_{\text{HL}}$ are a tensor product of up-spin position
eigenstates and down-spin orbitals,%
\begin{equation}
\left\vert \mathbf{R}\right\rangle \otimes\bigwedge\limits_{n=1,\cdots
,N}\left\vert f_{j_{n}}(\mathbf{R})\right\rangle ,
\end{equation}
with corresponding eigenvalue%
\begin{equation}
\sum_{n=1,\cdots,N}E_{j_{n}}(\mathbf{R}).
\end{equation}

We can use the eigenstates of $H_{\text{HL}}$ to perform a basis decomposition
of the ground state $\left\vert \Psi_{N,N}^{0}\right\rangle $. \ For any given
$\mathbf{R}$ and orbital indices $j_{1},\cdots,j_{N},$ we write the inner
product with the ground state as%
\begin{equation}
F_{\mathbf{R,}\left\{  j_{n}\right\}  }=\left[  \left\langle \mathbf{R}%
\right\vert \otimes\bigwedge\limits_{n=1,\cdots,N}\left\langle f_{j_{n}%
}(\mathbf{R})\right\vert \right]  \left\vert \Psi_{N,N}^{0}\right\rangle ,
\end{equation}
with normalization%
\begin{equation}
\int d^{dN}\mathbf{R}\sum_{\left\{  j_{n}\right\}  }\left\vert F_{\mathbf{R,}%
\left\{  j_{n}\right\}  }\right\vert ^{2}=1.
\end{equation}
We find that%
\begin{align}
E_{N,N}^{0}  &  =\left\langle \Psi_{N,N}^{0}\right\vert H\left\vert \Psi
_{N,N}^{0}\right\rangle =\left\langle \Psi_{N,N}^{0}\right\vert H_{\text{HL}%
}\left\vert \Psi_{N,N}^{0}\right\rangle \nonumber\\
&  =\int d^{dN}\mathbf{R}\sum_{\left\{  j_{n}\right\}  }\left\vert
F_{\mathbf{R,}\left\{  j_{n}\right\}  }\right\vert ^{2}\sum_{n=1,\cdots
,N}E_{j_{n}}(\mathbf{R}). \label{energy}%
\end{align}
The coefficients $\left\vert F_{\mathbf{R,}\left\{  j_{n}\right\}
}\right\vert ^{2}$ define a normalized probability distribution. \ Weighted by
this probability distribution, $E_{N,N}^{0}$ is the average over all
$\mathbf{R}$ and $\left\{  j_{n}\right\}  $ of the orbital energy sums%
\begin{equation}
\sum_{n=1,\cdots,N}E_{j_{n}}(\mathbf{R}).
\end{equation}

The energy constraint Eq.~(\ref{energy}) is an exact relation satisfied by the
ground state energy $E_{N,N}^{0}$ of the original Hamiltonian $H$. \ Since the
phase of $F_{\mathbf{R,}\left\{  j_{n}\right\}  }$ is irrelevant, solving this
constraint is qualitatively easier than solving for the ground state
wavefunction $\left\vert \Psi_{N,N}^{0}\right\rangle $. \ In this respect
Eq.~(\ref{energy}) can be regarded as the starting point for a generalized
density functional approach. \ However instead of single-particle densities we
work with many-body densities given by the coefficients $\left\vert
F_{\mathbf{R,}\left\{  j_{n}\right\}  }\right\vert ^{2}$.

\section{$H$ and $H_{\text{HL}}$ on the lattice}

Throughout this section we discuss Hamiltonian lattice regularization for
interacting two-component fermions with short-range interactions. \ Further
details for both Hamiltonian and Euclidean lattice formulations can be found
in
\cite{Chen:2003vy,Lee:2004qd,Lee:2004si,Wingate:2005xy,Bulgac:2005a,Lee:2005is,Lee:2005it,Lee:2005fk,Burovski:2006a,Burovski:2006b,Lee:2006hr,Lee:2007eu,Lee:2007jd}%
.

\subsection{Attractive Hubbard model}

On a $d$-dimensional spatial lattice we can write $H$ as%
\begin{equation}
H=K_{\uparrow}+K_{\downarrow}+V, \label{H}%
\end{equation}
where%
\begin{equation}
K_{\uparrow}=\frac{1}{2m}\sum_{\vec{r},l}\left[  2a_{\uparrow}^{\dagger}%
(\vec{r})a_{\uparrow}(\vec{r})-a_{\uparrow}^{\dagger}(\vec{r})a_{\uparrow
}(\vec{r}+\hat{l})-a_{\uparrow}^{\dagger}(\vec{r})a_{\uparrow}(\vec{r}-\hat
{l})\right]  ,
\end{equation}%
\begin{equation}
K_{\downarrow}=\frac{1}{2m}\sum_{\vec{r},l}\left[  2a_{\downarrow}^{\dagger
}(\vec{r})a_{\downarrow}(\vec{r})-a_{\downarrow}^{\dagger}(\vec{r}%
)a_{\downarrow}(\vec{r}+\hat{l})-a_{\downarrow}^{\dagger}(\vec{r}%
)a_{\downarrow}(\vec{r}-\hat{l})\right]  ,
\end{equation}
and%
\begin{equation}
V=C\sum_{\vec{r}}a_{\downarrow}^{\dagger}(\vec{r})a_{\uparrow}^{\dagger}%
(\vec{r})a_{\uparrow}(\vec{r})a_{\downarrow}(\vec{r}).
\end{equation}
Here $\vec{r}$ is an integer-valued $d$-dimensional spatial lattice vector,
and $\hat{l}=\hat{1},...,\hat{d}$ are lattice unit vectors in each of the
spatial directions. \ We write $m$ for the fermion mass and $C$ for the
coupling constant. \ Throughout we use dimensionless parameters and operators,
which correspond with physical values multiplied by the appropriate power of
the spatial lattice spacing $a$. \ This lattice model is the same as the
$d$-dimensional attractive Hubbard model. \ The Hamiltonian for the Hubbard
model is usually written as
\begin{equation}
H_{\text{Hubbard}}=-t\sum_{\vec{r},l,\sigma=\uparrow,\downarrow}\left[
a_{\sigma}^{\dagger}(\vec{r})a_{\sigma}(\vec{r}+\hat{l})+a_{\sigma}^{\dagger
}(\vec{r})a_{\sigma}(\vec{r}-\hat{l})\right]  +U\sum_{\vec{r}}a_{\downarrow
}^{\dagger}(\vec{r})a_{\uparrow}^{\dagger}(\vec{r})a_{\uparrow}(\vec
{r})a_{\downarrow}(\vec{r}).
\end{equation}
Therefore in $d$ dimensions,%
\begin{equation}
H_{\text{Hubbard}}=H+\left(  N_{\uparrow}+N_{\downarrow}\right)  \frac{d}{m}%
\end{equation}
and%
\begin{equation}
U=C,\quad t=\frac{1}{2m}.
\end{equation}

We define the heavy-light Hamiltonian on the lattice as%
\begin{equation}
H_{\text{HL}}=H-K_{\uparrow}+K_{\downarrow}=2K_{\downarrow}+V.
\end{equation}
As in the continuum case the change from $H$ to $H_{\text{HL}}$ can be viewed
as altering the masses of the spins. \ In $H$ we have $m_{\uparrow
}=m_{\downarrow}=m$, while in $H_{\text{HL}}$ we have $m_{\uparrow}%
=\infty,m_{\downarrow}=m/2$. \ More generally we can define%
\begin{equation}
H(\theta)=H+\left(  -K_{\uparrow}+K_{\downarrow}\right)  \theta=(1-\theta
)K_{\uparrow}+(1+\theta)K_{\downarrow}+V,
\end{equation}
which interpolates between $H$ at $\theta=0$ and $H_{\text{HL}}$ at $\theta
=1$. \ In terms of the spin-dependent masses $m_{\uparrow}$ and $m_{\downarrow
}$,%
\begin{equation}
\theta=\frac{m_{\uparrow}-m_{\downarrow}}{m_{\uparrow}+m_{\downarrow}}.
\end{equation}

\subsection{Spectrum of the four-body system in three dimensions}

To gain some intuition for the low-energy spectrum of $H(\theta)$, we consider
the unpolarized four-body system, $N_{\uparrow}=N_{\downarrow}=2$, in three
dimensions. \ We tune the coupling constant $C$ to the unitarity limit where
the scattering length is infinite. \ This procedure uses L\"{u}scher's
finite-volume scattering formula and is discussed in the appendix. \ In
Fig.~\ref{theta_dependence} we show the low-energy spectrum of $H(\theta)$ for
the four-body system at total momentum $\vec{P}=0$ on a $4^{3}$ lattice. \ The
energies are computed using the Lanczos method \cite{Lanczos:1950} and written
as a fraction of the ground state energy $E_{2,2}^{0,\text{free}}$ for the
non-interacting system. \ The concentric circles indicate exact degeneracies
as well as approximate degeneracies too close to visually distinguish in the plot.%

\begin{figure}
[ptb]
\begin{center}
\includegraphics[
height=2.4241in,
width=3.3875in
]%
{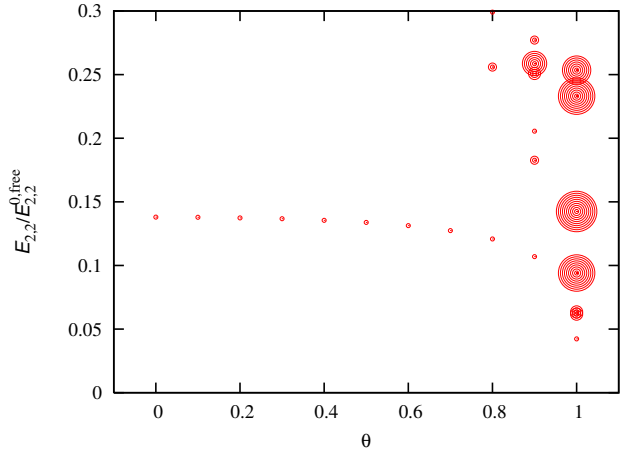}%
\caption{Low-energy spectrum of $H(\theta)$ at total momentum $\vec{P}=0$ for
the four-body system on a $4^{3}$ lattice \ The energies are written as a
fraction of the ground state energy $E_{2,2}^{0,\text{free}}$ for the
non-interacting system. \ The coupling constant $C$ is tuned to the unitarity
point.}%
\label{theta_dependence}%
\end{center}
\end{figure}
\begin{figure}
[ptbptb]
\begin{center}
\includegraphics[
height=2.3177in,
width=2.3177in
]%
{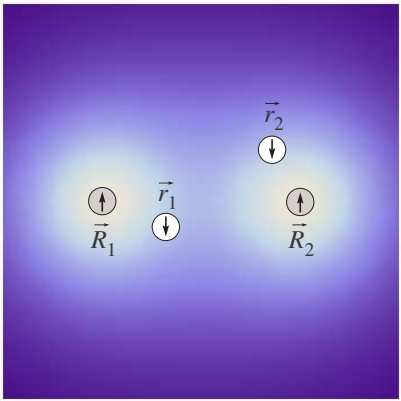}%
\caption{Sketch of the one-body density distribution of down spins in the
lowest orbitals for a fixed configuration of up spins.}%
\label{lowestfilling}%
\end{center}
\end{figure}

The ground state energy remains relatively constant for $\theta<0.6$. \ For
larger $\theta$ the ground state energy decreases more substantially while the
density of low-energy states increases. \ The data at $\theta=1$ corresponds
with the spectrum of the heavy-light Hamiltonian $H_{\text{HL}}$. \ From
Fig.~\ref{theta_dependence} we count 35 energy states at $\theta=1$ with
$E_{2,2}/E_{2,2}^{0,\text{free}}$ less than $0.25$. \ These lowest $35$ energy
states correspond with the $35$ independent states obtained by placing down
spins in the lowest two energy orbitals for each possible up-spin
configuration $\mathbf{R=\{}\vec{R}_{1},\vec{R}_{2}\}$, and then projecting
onto $\vec{P}=0$. \ A\ sketch of the one-body density distribution of down
spins in the lowest orbitals for a fixed configuration of up spins is shown in
Fig.~\ref{lowestfilling}.%

\section{Lowest filling approximation}

\subsection{Lowest orbital filling and effective distribution $\left\vert
F_{\mathbf{R}}\right\vert ^{2}$}

On the lattice we write a discrete sum over $\mathbf{R}$ rather than a
continuous integral. \ Therefore Eq.~(\ref{energy}) becomes%
\begin{equation}
E_{N,N}^{0}=\sum_{\mathbf{R}}\sum_{\left\{  j_{n}\right\}  }\left\vert
F_{\mathbf{R,}\left\{  j_{n}\right\}  }\right\vert ^{2}\sum_{n=1,\cdots
,N}E_{j_{n}}(\mathbf{R}). \label{energy2}%
\end{equation}
In our example of the four-body system in three dimensions at unitarity, we
found that for each $\mathbf{R}$ the sum%
\begin{equation}
\sum_{n=1,2}E_{j_{n}}(\mathbf{R})
\end{equation}
is numerically close to $E_{2,2}^{0}$ only when the down spins occupy the
lowest orbitals, $\left\{  j_{n}\right\}  =\{1,2\}$. \ Given that the weighted
average
\begin{equation}
\sum_{\mathbf{R}}\sum_{\left\{  j_{n}\right\}  }\left\vert F_{\mathbf{R,}%
\left\{  j_{n}\right\}  }\right\vert ^{2}\sum_{n=1,2}E_{j_{n}}(\mathbf{R})
\end{equation}
equals $E_{2,2}^{0}$, we conclude that $\left\vert F_{\mathbf{R,}\left\{
j_{n}\right\}  }\right\vert ^{2}$ is dominated by the set of lowest orbitals
$\left\{  j_{n}\right\}  =\{1,2\}$ for each $\mathbf{R}$. \ This suggests a
general approximation scheme which keeps only the lowest orbitals.

As in our four-body example at unitarity, let us assume that most of the
weight of the normalized distribution $\left\vert F_{\mathbf{R,}\left\{
j_{n}\right\}  }\right\vert ^{2}$ lies in orbital sets $\left\{
j_{n}\right\}  $ where all orbitals lie near the bottom $N$ in energy. \ Let%
\begin{equation}
F_{\mathbf{R}}=\sqrt{\sum_{\left\{  j_{n}\right\}  }\left\vert F_{\mathbf{R,}%
\left\{  j_{n}\right\}  }\right\vert ^{2}}.
\end{equation}
Then%
\begin{equation}
E_{N,N}^{0}=\sum_{\mathbf{R}}\sum_{\left\{  j_{n}\right\}  }\left\vert
F_{\mathbf{R,}\left\{  j_{n}\right\}  }\right\vert ^{2}\sum_{n=1,\cdots
,N}E_{j_{n}}(\mathbf{R})\approx\sum_{\mathbf{R}}\left\vert F_{\mathbf{R}%
}\right\vert ^{2}\sum_{j=1,\cdots,N}E_{j}(\mathbf{R}).
\end{equation}
This approximation uses an effective distribution with weight $\left\vert
F_{\mathbf{R}}\right\vert ^{2}$ which is nonzero only for $\left\{
j_{n}\right\}  =\left\{  1,\cdots,N\right\}  $. \ We refer to this as the
lowest filling approximation.

\subsection{$N$-body fixed-point densities}

In the lowest filling approximation we are restricted to states of the general
form%
\begin{equation}
\left\vert \Phi_{N,N}\right\rangle =\sum_{\mathbf{R}}F_{\mathbf{R}}\left[
\left\vert \mathbf{R}\right\rangle \otimes\bigwedge\limits_{j=1,\cdots
,N}\left\vert f_{j}(\mathbf{R})\right\rangle \right]  , \label{lowest}%
\end{equation}%
\begin{equation}
\sum_{\mathbf{R}}\left\vert F_{\mathbf{R}}\right\vert ^{2}=1.
\end{equation}
In general any $\left\vert \Phi_{N,N}\right\rangle $ of this form will not
satisfy the $Z_{2}$-invariance condition $U\left\vert \Phi_{N,N}\right\rangle
=\left\vert \Phi_{N,N}\right\rangle $ exactly. \ Therefore we search for a
weaker symmetry constraint that follows from $Z_{2}$ invariance but can be
satisfied exactly in the lowest filling approximation. \ In the following we
design a constraint which is exactly solvable in the lowest filling
approximation, with a unique solution $\left\vert F_{\mathbf{R}}\right\vert
^{2}$ for each $\mathbf{R}$.

Let us define the up-spin and down-spin particle densities,%
\begin{equation}
\rho_{\uparrow}(\vec{r})=a_{\uparrow}^{\dagger}(\vec{r})a_{\uparrow}(\vec{r}),
\end{equation}%
\begin{equation}
\rho_{\downarrow}(\vec{r})=a_{\downarrow}^{\dagger}(\vec{r})a_{\downarrow
}(\vec{r}).
\end{equation}
Let $G\left[  \rho_{\uparrow},\rho_{\downarrow}\right]  $ be any functional
involving the particle densities. \ Then any state $\left\vert \Phi
_{N,N}\right\rangle $ satisfying the invariance condition $U\left\vert
\Phi_{N,N}\right\rangle =\left\vert \Phi_{N,N}\right\rangle $ also satisfies%
\begin{equation}
\left\langle \Phi_{N,N}\right\vert :G\left[  \rho_{\uparrow},\rho_{\downarrow
}\right]  :\left\vert \Phi_{N,N}\right\rangle =\left\langle \Phi
_{N,N}\right\vert :G\left[  \rho_{\downarrow},\rho_{\uparrow}\right]
:\left\vert \Phi_{N,N}\right\rangle .
\end{equation}
In particular we have%
\begin{equation}
\left\langle \Phi_{N,N}\right\vert :\rho_{\uparrow}(\vec{r}_{1})\times
\cdots\times\rho_{\uparrow}(\vec{r}_{N}):\left\vert \Phi_{N,N}\right\rangle
=\left\langle \Phi_{N,N}\right\vert :\rho_{\downarrow}(\vec{r}_{1}%
)\times\cdots\times\rho_{\downarrow}(\vec{r}_{N}):\left\vert \Phi
_{N,N}\right\rangle . \label{densityN}%
\end{equation}
The $:$ symbols denote normal ordering. \ Upon summation this constraint
implies analogous relations for the $j$-body densities for each $j<N$,%
\begin{equation}
\left\langle \Phi_{N,N}\right\vert :\rho_{\uparrow}(\vec{r}_{1})\times
\cdots\times\rho_{\uparrow}(\vec{r}_{j}):\left\vert \Phi_{N,N}\right\rangle
=\left\langle \Phi_{N,N}\right\vert :\rho_{\downarrow}(\vec{r}_{1}%
)\times\cdots\times\rho_{\downarrow}(\vec{r}_{j}):\left\vert \Phi
_{N,N}\right\rangle .
\end{equation}
We can write the $N$-body up-spin densities in terms of $F_{\mathbf{R}}$,%
\begin{align}
\left\langle \Phi_{N,N}\right\vert  &  :\rho_{\uparrow}(\vec{R}_{1}%
)\times\cdots\times\rho_{\uparrow}(\vec{R}_{N}):\left\vert \Phi_{N,N}%
\right\rangle \nonumber\\
&  =\left\langle \Phi_{N,N}\right\vert a_{\uparrow}^{\dag}(\vec{R}_{N}%
)\times\cdots\times a_{\uparrow}^{\dag}(\vec{R}_{1})a_{\uparrow}(\vec{R}%
_{1})\times\cdots\times a_{\uparrow}(\vec{R}_{N})\left\vert \Phi
_{N,N}\right\rangle =N!\left\vert F_{\mathbf{R}}\right\vert ^{2}.
\end{align}
We can write the $N$-body down-spin densities in terms of $F_{\mathbf{R}}$ and
the inner product $\left\langle \mathbf{r}\right\vert \bigwedge
\limits_{j=1,\cdots,N}\left\vert f_{j}(\mathbf{R})\right\rangle $,%
\begin{equation}
\left\langle \Phi_{N,N}\right\vert :\rho_{\downarrow}(\vec{r}_{1})\times
\cdots\times\rho_{\downarrow}(\vec{r}_{N}):\left\vert \Phi_{N,N}\right\rangle
=N!\sum_{\mathbf{R}}\left\vert \left\langle \mathbf{r}\right\vert
\bigwedge\limits_{j=1,\cdots,N}\left\vert f_{j}(\mathbf{R})\right\rangle
\right\vert ^{2}\times\left\vert F_{\mathbf{R}}\right\vert ^{2}.
\end{equation}
If the $N$-body densities for up spins and down spins are equal then%
\begin{equation}
\sum_{\mathbf{R}}\left\vert \left\langle \mathbf{r}\right\vert \bigwedge
\limits_{j=1,\cdots,N}\left\vert f_{j}(\mathbf{R})\right\rangle \right\vert
^{2}\mathbf{\times}\left\vert F_{\mathbf{R}}\right\vert ^{2}=\left\vert
F_{\mathbf{r}}\right\vert ^{2}. \label{fixedpoint}%
\end{equation}
We refer to any $\left\vert F_{\mathbf{R}}\right\vert ^{2}$ which satisfies
this relation an $N$-body fixed-point density. \ The existence and uniqueness
of fixed-point solutions $\left\vert F_{\mathbf{R}}\right\vert ^{2}$ will be
proved in the next section.

At this point we comment on the clustering instability of $H_{\text{HL}}.$\ In
two dimensions when any two of the up-spin locations $\vec{R}_{i}$ and
$\vec{R}_{j}$ come close together, the lowest heavy-light orbital energy in
physical units has a divergence proportional to $a^{-1}$, where $a$ is the
lattice spacing. \ In three dimensions the divergence is $a^{-2}$. \ However
any $N$-body fixed-point distribution $\left\vert F_{\mathbf{R}}\right\vert
^{2}$ must vanish as $\left\vert \vec{R}_{i}-\vec{R}_{j}\right\vert ^{2}$ due
to antisymmetry with respect to exchange of $\vec{r}_{i}$ and $\vec{r}_{j}$
in
\begin{equation}
\left\langle \mathbf{r}\right\vert \bigwedge\limits_{j=1,\cdots,N}\left\vert
f_{j}(\mathbf{R})\right\rangle .
\end{equation}
This is enough to remove the clustering instability for $d\leq3$. \ Therefore
the continuum limit is well defined for the lowest filling approximation to
the ground state energy,%
\begin{equation}
E_{N,N}^{0}\approx\sum_{\mathbf{R}}\left\vert F_{\mathbf{R}}\right\vert
^{2}\sum_{j=1,\cdots,N}E_{j}(\mathbf{R}),
\end{equation}
where $\left\vert F_{\mathbf{R}}\right\vert ^{2}$ is an $N$-body fixed-point density.

\subsection{Lowest filling approximation for the four-body system in three
dimensions}

We test how well the lowest filling approximation works for the unpolarized
four-body system in three dimensions. \ Again the coupling is tuned to the
unitarity point and the lattice volume is $4^{3}$. \ Once we project onto the
total momentum $\vec{P}=0$ subspace, there are $35$ independent configurations
$\mathbf{R}$ for the up spins. \ Similarly there are $35$ independent
configurations for the down spin positions $\mathbf{r}$. \ We construct a
$35\times35$ matrix $M(\mathbf{r,R})$ with elements%
\begin{equation}
M(\mathbf{r,R})=\left\vert \left\langle \mathbf{r}\right\vert \bigwedge
\limits_{j=1,2}\left\vert f_{j}(\mathbf{R})\right\rangle \right\vert ^{2},
\end{equation}
where $\left\vert f_{1}(\mathbf{R})\right\rangle ,\left\vert f_{2}%
(\mathbf{R})\right\rangle $ are the lowest orbitals for $\mathbf{R}$. \ We
then solve for the $N$-body fixed-point density $\left\vert F_{\mathbf{R}%
}\right\vert ^{2}$ satisfying%
\begin{equation}
\sum_{\mathbf{R}}M(\mathbf{r,R})\left\vert F_{\mathbf{R}}\right\vert
^{2}=\left\vert F_{\mathbf{r}}\right\vert ^{2}.
\end{equation}

\begin{figure}
[ptb]
\begin{center}
\includegraphics[
height=2.4388in,
width=4.2272in
]%
{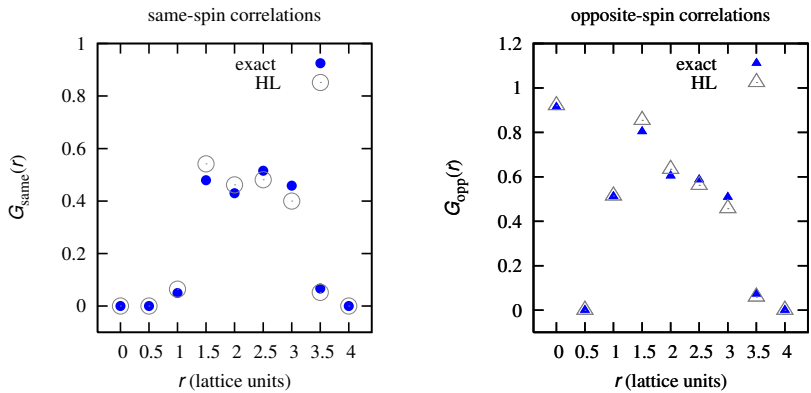}%
\caption{Comparison of exact and heavy-light results for the same-spin and
opposite-spin correlation functions. \ The data is binned by radial distance
in lattice units rounded to the nearest half integer.}%
\label{twobodydensity}%
\end{center}
\end{figure}

With $\left\vert F_{\mathbf{R}}\right\vert ^{2}$ the ground state energy can
be calculated using%
\begin{equation}
E_{2,2}^{0}\approx\sum_{\mathbf{R}}\left\vert F_{\mathbf{R}}\right\vert
^{2}\sum_{j=1,2}E_{j}(\mathbf{R}).
\end{equation}
We can also compute two-body correlation functions. \ The same-spin
correlation function in the lowest filling approximation is%
\begin{align}
G_{\text{same}}(\vec{r})  &  =\sum_{\vec{r}^{\,\prime}}\left\langle \Psi
_{2,2}^{0}\right\vert :\rho_{\downarrow}(\vec{r}+\vec{r}^{\,\prime}%
)\rho_{\downarrow}(\vec{r}^{\,\prime}):\left\vert \Psi_{2,2}^{0}\right\rangle
\nonumber\\
&  =\sum_{\vec{r}^{\,\prime}}\left\langle \Psi_{2,2}^{0}\right\vert
:\rho_{\uparrow}(\vec{r}+\vec{r}^{\,\prime})\rho_{\uparrow}(\vec{r}^{\,\prime
}):\left\vert \Psi_{2,2}^{0}\right\rangle \nonumber\\
&  \approx\sum_{\mathbf{R}}\left(  \delta_{\vec{R}_{1},\vec{r}+\vec{R}_{2}%
}+\delta_{\vec{R}_{2},\vec{r}+\vec{R}_{1}}\right)  \left\vert F_{\mathbf{R}%
}\right\vert ^{2}.
\end{align}
In our dimensionless lattice normalization the same-spin correlation function
summed over $\vec{r}$ equals $2.$ \ Also the opposite-spin correlation
function is given by%
\begin{align}
G_{\text{opp}}(\vec{r})  &  =\sum_{\vec{r}^{\,\prime}}\left\langle \Psi
_{2,2}^{0}\right\vert :\rho_{\uparrow}(\vec{r}+\vec{r}^{\,\prime}%
)\rho_{\downarrow}(\vec{r}^{\,\prime}):\left\vert \Psi_{2,2}^{0}\right\rangle
\nonumber\\
&  =\sum_{\vec{r}^{\,\prime}}\left\langle \Psi_{2,2}^{0}\right\vert
:\rho_{\downarrow}(\vec{r}+\vec{r}^{\,\prime})\rho_{\uparrow}(\vec
{r}^{\,\prime}):\left\vert \Psi_{2,2}^{0}\right\rangle \nonumber\\
&  \approx\sum_{\mathbf{R}}\sum_{j=1,2}\left[  \left\vert f_{j}(\vec{r}%
+\vec{R}_{1},\mathbf{R})\right\vert ^{2}+\left\vert f_{j}(\vec{r}+\vec{R}%
_{2},\mathbf{R})\right\vert ^{2}\right]  \left\vert F_{\mathbf{R}}\right\vert
^{2}.
\end{align}
The opposite-spin correlation function summed over $\vec{r}$ equals $4$.
\ Fig.~\ref{twobodydensity} shows a comparison of exact results and
heavy-light results in the lowest filling approximation for the same-spin and
opposite-spin correlation functions. \ The exact results are Lanczos
calculations for the ground state of $H$. \ The data is binned by radial
distance $r=\left\vert \vec{r}\right\vert $ in lattice units rounded to the
nearest half integer.%
The agreement between exact and heavy-light results for the correlation
functions is rather good. \ The agreement for the ground state energy is also
quite good. \ The exact value for the ratio $E_{2,2}^{0}/E_{2,2}%
^{0\text{,free}}$ is $0.138$ while the heavy-light result gives $0.128$.
\ This level of quantitative agreement may seem surprising given the
simplicity of the lowest filling approximation to the heavy-light ansatz.
\ The method simply patches together a collection of single-particle orbitals
using the $N$-body fixed-point constraint in Eq.~(\ref{fixedpoint}). \ However
the method appears to accurately describe the competition between the
short-range attractive interaction and Fermi repulsion. \ We now check whether
the agreement is also good for different interaction strengths.

The ratio $E_{N,N}^{0}/E_{N,N}^{0\text{,free}}$ tends towards negative
infinity at strong attractive coupling and is therefore somewhat unwieldy to
plot over a large range of coupling strengths. \ Let us define a different
dimensionless ratio which is convenient for the attractive Hubbard model at
arbitrary coupling and arbitrary dimensions. \ At fixed lattice volume $L^{d}$
we define the ratio%
\begin{equation}
e_{N,N}=\frac{E_{N,N}^{0}-E_{N,N}^{0,\text{free}}}{N\left\vert E_{1,1}%
^{0}\right\vert +E_{N,N}^{0,\text{free}}},
\end{equation}
where $E_{1,1}^{0}$ is the ground state energy for the two-body dimer in the
same lattice volume $L^{d}$. \ In the limit of weak attractive coupling,
$e_{N,N}$ tends towards $0$ from below. \ In this case the physics is
dominated by Fermi repulsion, and the ground state is similar to that of a
free Fermi gas. \ In the limit of strong attractive coupling, $e_{N,N}$ tends
towards $-1$ from above. \ Here the ground state consists of tightly-bound
dimers with only weak interactions between dimers. \ For $e_{N,N}$ in the
midrange between $0$ and $-1$, the competition between attractive interaction
and Fermi repulsion is stalemated to some intermediate balance point. \ The
unitarity limit in three dimensions fits this category. \ In the limit of
large $N$ at unitarity the $N\left\vert E_{1,1}^{0}\right\vert $ term can be
neglected relative to $E_{N,N}^{0,\text{free}}$. \ In this case%
\begin{equation}
e_{N,N}\rightarrow\frac{E_{N,N}^{0}-E_{N,N}^{0,\text{free}}}{E_{N,N}%
^{0,\text{free}}}=E_{N,N}^{0}/E_{N,N}^{0\text{,free}}-1.
\end{equation}

\begin{figure}
[ptb]
\begin{center}
\includegraphics[
height=2.6057in,
width=4.4443in
]%
{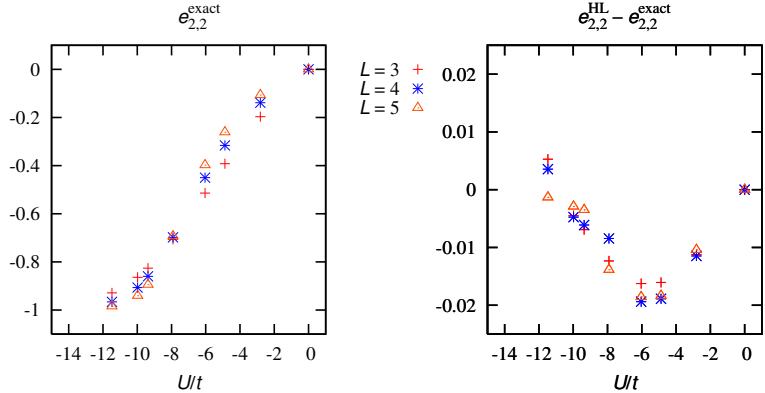}%
\caption{Comparison of four-body exact Lanczos results, $e_{2,2}%
^{\text{exact}}$, and heavy-light results, $e_{2,2}^{\text{HL}}$, for
$L=3,4,5$ lattices in three dimensions plotted versus coupling $U/t=2mC.$}%
\label{threed_22}%
\end{center}
\end{figure}

In Fig.~\ref{threed_22} we show results for $e_{2,2}$ for lattice lengths
$L=3,4,5$ and various couplings. \ We express the coupling as a ratio of
Hubbard model parameters%
\begin{equation}
U/t=2mC.
\end{equation}
The plot on the left shows exact results computed using the Lanczos method.
\ \ The unitarity point corresponds with $U/t=-7.914$, and the signal of scale
invariance at unitarity can be seen by the agreement in $e_{2,2}$ for
different $L$. \ The plot on the right shows the difference between
heavy-light results in the lowest filling approximation and exact Lanczos
results.
From weak coupling at $e_{2,2}\approx0$ to strong coupling at $e_{2,2}%
\approx-1$, the error of the heavy-light calculation is bounded by $0.02$.
\ The method appears to accurately describe the crossover from four
weakly-interacting fermions to a condensed pair of bosonic dimers.

\section{Markov chain for $N$-body fixed-point densities}

In this section we design a Markov chain process which generates the $N$-body
fixed-point density $\left\vert F_{\mathbf{R}}\right\vert ^{2}$ that solves
the constraint in Eq.~(\ref{fixedpoint}). \ Our construction establishes both
existence and uniqueness. \ Let us define
\begin{equation}
\Upsilon(\mathbf{r,R)=}\left\vert \left\langle \mathbf{r}\right\vert
\bigwedge\limits_{j=1,\cdots,N}\left\vert f_{j}(\mathbf{R})\right\rangle
\right\vert ^{2}. \label{upsilon}%
\end{equation}
We note that $\Upsilon(\mathbf{r,R)}\geq0$ for every pair $\mathbf{r,R}$.
\ Also for each $\mathbf{R}$,%
\begin{equation}
\sum_{\mathbf{r}}\Upsilon(\mathbf{r,R)}=1.
\end{equation}
We define a Markov chain
\begin{equation}
\mathbf{R}^{(0)}\rightarrow\mathbf{R}^{(1)}\rightarrow\cdots\rightarrow
\mathbf{R}^{(k)}\rightarrow\mathbf{R}^{(k+1)}\rightarrow\cdots
\end{equation}
with transition probability%
\begin{equation}
p\left(  \mathbf{R}^{(k+1)}\right\vert \left.  \mathbf{R}^{(k)}\right)
=\Upsilon(\mathbf{R}^{(k+1)},\mathbf{R}^{(k)}\mathbf{).}%
\end{equation}

Each state in the Markov chain is a set of $N$ distinct $d$-dimensional
vectors on an $L^{d}$ lattice. \ We note that
\begin{equation}
\Upsilon(\mathbf{R}^{(k+1)},\mathbf{R}^{(k)}\mathbf{)}=0
\end{equation}
if and only if $\left\vert \mathbf{R}^{(k+1)}\right\rangle $ and
$\bigwedge\limits_{j=1,\cdots,N}\left\vert f_{j}(\mathbf{R}^{(k)}%
)\right\rangle $ are exactly orthogonal. \ Since the state space is finite the
possibility of accidental orthogonality between $\left\vert \mathbf{R}%
^{(k+1)}\right\rangle $ and $\bigwedge\limits_{j=1,\cdots,N}\left\vert
f_{j}(\mathbf{R}^{(k)})\right\rangle $ is a set of measure zero and can
avoided by an arbitrarily small change in the coupling $C$. \ However it is
also possible that the orthogonality arises from some mismatch of exactly
conserved quantum numbers. \ These quantum numbers would be associated with
some symmetry subgroup of the lattice shared by $\mathbf{R}^{(k+1)}$ and
$\mathbf{R}^{(k)}$. \ It might be a reflection symmetry, rotational symmetry,
translational symmetry, or some combination of each. \ However for large $L$
the relative proportion of such symmetric configurations $\mathbf{R}^{(k+1)}$
and $\mathbf{R}^{(k)}$ is suppressed by powers of $L$ and therefore
exceedingly rare. \ We let $\epsilon$ be a small positive number and define a
modified chain with transition probability matrix%
\begin{equation}
p\left(  \mathbf{R}^{(k+1)}\right\vert \left.  \mathbf{R}^{(k)}\right)
=\Upsilon_{\epsilon}(\mathbf{R}^{(k+1)},\mathbf{R}^{(k)}\mathbf{),}%
\end{equation}%
\begin{equation}
\Upsilon_{\epsilon}(\mathbf{r,R})=\frac{\max\left[  \Upsilon(\mathbf{r,R}%
),\epsilon\right]  }{\sum\limits_{\mathbf{r}}\max\left[  \Upsilon
(\mathbf{r,R}),\epsilon\right]  }. \label{upsilon_eps}%
\end{equation}
For any $\mathbf{R}^{\prime}$ and $\mathbf{R}^{\prime\prime}$ there exists
some chain of finite length from $\mathbf{R}^{\prime}$ to $\mathbf{R}%
^{\prime\prime}$ such that the transition probability at each step is nonzero.
\ In fact we can get there in only one step. \ Hence the Markov chain is
ergodic and there exists a unique invariant distribution, call it $\left\vert
F_{\mathbf{R}}\right\vert ^{2}$, such that%
\begin{equation}
\sum_{\mathbf{R}}\Upsilon_{\epsilon}(\mathbf{R}^{\prime}\mathbf{,R)\times
}\left\vert F_{\mathbf{R}}\right\vert ^{2}=\left\vert F_{\mathbf{R}^{\prime}%
}\right\vert ^{2}.
\end{equation}
We take the limit as $\epsilon\rightarrow0$ and obtain $\left\vert
F_{\mathbf{R}}\right\vert ^{2}$ as the unique $N$-body fixed-point density.

This fixed-point Markov chain shares some features with iterative solutions of
the Kohn-Sham equations in density functional theory \cite{KohnSham:1965}.
\ Both involve finding the lowest orbitals of a Schr\"{o}dinger potential
which in turn depends on the particle densities. \ However there are important
differences between this method and the Kohn-Sham equations. \ The fixed-point
Markov chain presented above solves for $N$-body densities rather than
one-body densities. \ Also in contrast with density functional theory, this
method yields an ab initio solution for the uniform system. \ We discuss
possible extensions to non-uniform systems in the discussion section.

As a precise test of the Markov chain algorithm for $\left\vert F_{\mathbf{R}%
}\right\vert ^{2}$ we compare with heavy-light calculations presented earlier
for the unpolarized four-body system at unitarity on a $4^{3}$ lattice. \ For
this calculation and all other heavy-light calculations presented in the
following we generate the fixed-point $\left\vert F_{\mathbf{R}}\right\vert
^{2}$ using 4000 steps of the Markov chain%
\begin{equation}
\mathbf{R}^{(0)}\rightarrow\mathbf{R}^{(1)}\rightarrow\cdots\rightarrow
\mathbf{R}^{(k)}\rightarrow\mathbf{R}^{(k+1)}\rightarrow\cdots
\end{equation}
with transition probability%
\begin{equation}
p\left(  \mathbf{R}^{(k+1)}\right\vert \left.  \mathbf{R}^{(k)}\right)
=\Upsilon(\mathbf{R}^{(k+1)},\mathbf{R}^{(k)}\mathbf{).}%
\end{equation}
Each step of the chain $\mathbf{R}^{(k)}\rightarrow\mathbf{R}^{(k+1)}$ is
produced using a Metropolis algorithm with $100$ updates per up-spin particle
location according to the probability distribution%
\begin{equation}
\Upsilon(\mathbf{R},\mathbf{R}^{(k)}\mathbf{)}=\left\vert \left\langle
\mathbf{R}\right\vert \bigwedge\limits_{j=1,\cdots,N}\left\vert f_{j}%
(\mathbf{R}^{(k)})\right\rangle \right\vert ^{2}.
\end{equation}
The entire calculation is repeated several times with different random number
seeds and the results are averaged. \ The standard deviation of the
distribution is used to determine the stochastic error of the average.

\begin{figure}
[ptb]
\begin{center}
\includegraphics[
height=2.4388in,
width=4.2272in
]%
{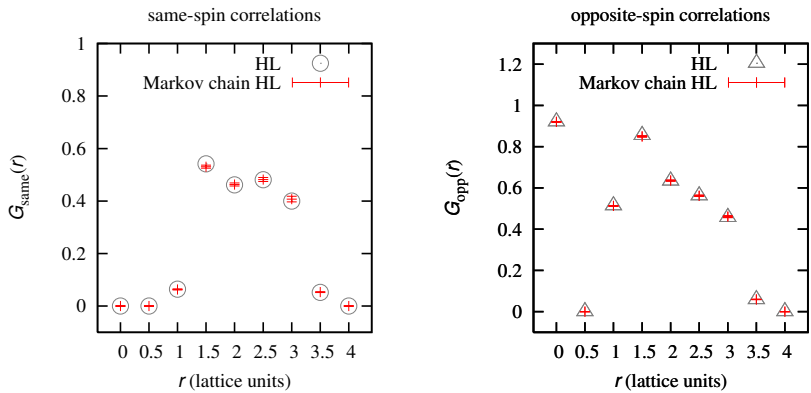}%
\caption{Comparison of Markov chain and direct heavy-light calculations for
the same-spin and opposite-spin correlation functions}%
\label{twobodydensity_markov}%
\end{center}
\end{figure}

In Fig.~\ref{twobodydensity_markov} we show a comparison of calculations for
the same-spin and opposite-spin correlations functions using the Markov chain
algorithm and the direct calculations shown previously in
Fig.~\ref{twobodydensity}. \ The error bars shown are estimated stochastic
errors.%
We see that the Markov chain algorithm reproduces the direct calculations for
the correlation functions with no detectable systematic error. \ The Markov
chain heavy-light result for $E_{2,2}^{0}/E_{2,2}^{0\text{,free}}$ is
$0.127(1)$, which agrees with the direct heavy-light calculation of $0.128$.
\ The advantage of the Markov chain algorithm is that the computational
scaling for large systems is very favorable, nearly linear in the lattice
volume and number of particles.

\section{Few-body systems in one and two dimensions}

In this section we compare heavy-light calculations for $e_{N,N}$ in the
lowest filling approximation with Lanczos calculations for few-body systems in
one and two spatial dimensions. \ The rapid computational scaling of the
Lanczos calculation, $L^{(2N-1)d}$, limits the system sizes for which the
comparison is possible. \ Let us first discuss the properties of bound dimers
in the 1D, 2D, and 3D attractive Hubbard models.

In one dimension at infinite volume the dimer has binding energy
\cite{Lee:2007jd}%
\begin{equation}
B_{2}=-E_{1,1}^{0}=\frac{mC^{2}}{4}+\cdots,
\end{equation}
and characteristic size%
\begin{equation}
\frac{1}{\sqrt{mB_{2}}}=\frac{2}{m\left\vert C\right\vert }+\cdots.\label{1d}%
\end{equation}
The ellipses denote lattice spacing corrections which become important when
the characteristic size is not much larger than one lattice spacing. \ In two
dimensions at infinite volume the dimer has binding energy \cite{Lee:2005xy}%
\begin{equation}
B_{2}=-E_{1,1}^{0}=\frac{3.24\pi^{2}}{m}\exp\left(  \frac{4\pi}{mC}\right)
+\cdots,
\end{equation}
and characteristic size%
\begin{equation}
\frac{1}{\sqrt{mB_{2}}}=\frac{1}{1.80\pi}\exp\left(  -\frac{2\pi}{mC}\right)
+\cdots.\label{2d}%
\end{equation}
In three dimensions at infinite volume and $mC<-3.957$, the dimer has binding
energy \cite{Lee:2005it,Bulgac:2005a,Burovski:2006a}%
\begin{equation}
B_{2}=-E_{1,1}^{0}=\frac{1}{ma_{\text{scatt}}^{2}}=\frac{16\pi^{2}}{m}\left(
\frac{1}{mC}+\frac{1}{3.957}\right)  ^{2}+\cdots,
\end{equation}
and characteristic size%
\begin{equation}
\frac{1}{\sqrt{mB_{2}}}=\frac{1}{4\pi\left(  \frac{1}{mC}+\frac{1}%
{3.957}\right)  }+\cdots.\label{3d}%
\end{equation}
In one and two dimensions the attractive Hubbard model has no nontrivial
scale-invariant point analogous to the unitarity point in three dimensions.

\begin{figure}
[ptb]
\begin{center}
\includegraphics[
height=2.6048in,
width=4.4443in
]%
{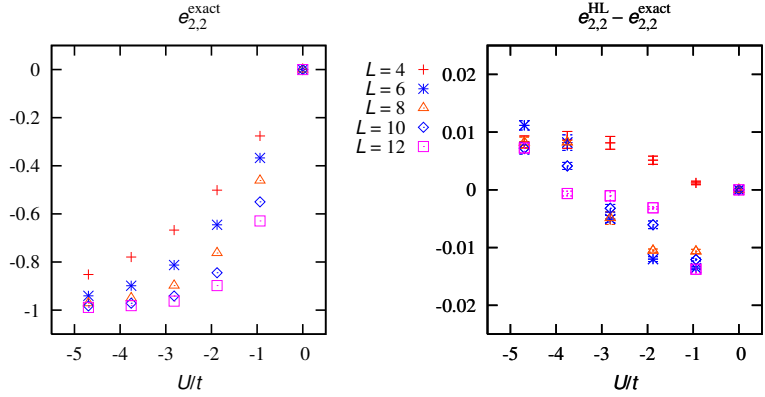}%
\caption{Comparison of four-body exact Lanczos results, $e_{2,2}%
^{\text{exact}}$, and heavy-light results, $e_{2,2}^{\text{HL}}$, for
$L=4,6,8,10,12$ lattices in one dimension plotted versus coupling $U/t=2mC.$}%
\label{oned_22}%
\end{center}
\end{figure}

\begin{figure}
[ptbptb]
\begin{center}
\includegraphics[
height=2.6048in,
width=4.4443in
]%
{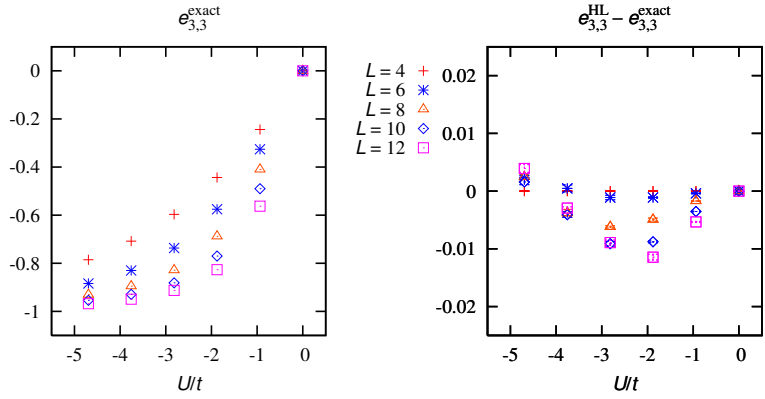}%
\caption{Comparison of six-body exact Lanczos results, $e_{3,3}^{\text{exact}%
}$, and heavy-light results, $e_{3,3}^{\text{HL}}$, for $L=4,6,8,10,12$
lattices in one dimension plotted versus coupling $U/t=2mC.$}%
\label{oned_33}%
\end{center}
\end{figure}

Fig.~\ref{oned_22} shows $e_{2,2}$ for the unpolarized four-body system in one
dimension.%
Fig.~\ref{oned_33} shows $e_{3,3}$ for the unpolarized six-body system in one
dimension.%
For both the four-body and six-body systems the error of the heavy-light
result is smaller than $0.015$. \ We note that the six-body heavy-light data
for $L=4$ is exact. \ This is because the system can be regarded as the ground
state of one up-spin hole and one down-spin hole. \ Similar to the unpolarized
two-particle system, the ground state of the unpolarized two-hole system
depends only on the reduced mass $\mu$,%
\begin{equation}
\frac{1}{\mu}=\frac{1}{m_{\uparrow}}+\frac{1}{m_{\downarrow}},
\end{equation}
which is unaltered in the heavy-light formalism.

Just as in the three-dimensional example at unitarity, the method appears to
accurately describe the crossover from fermions to condensed bosonic dimers in
one dimension. \ The difference between the ground state energy $E_{N,N}^{0}$
and $N$ times the dimer energy $E_{1,1}^{0}$ can be viewed as Fermi repulsion
of overlapping of dimer wavefunctions. \ In one dimension the dimer
wavefunction has an exponential tail proportional to%
\begin{equation}
\exp\left(  \frac{1}{2}mC\left\vert x\right\vert \right)  =\exp\left(
\frac{U}{4t}\left\vert x\right\vert \right)  .
\end{equation}
This simple picture is consistent with the exponential dependence of $e_{N,N}$
on $U/t$ seen in both the four-body and six-body results for large negative
$U/t$.

\begin{figure}
[ptb]
\begin{center}
\includegraphics[
height=2.6048in,
width=4.4443in
]%
{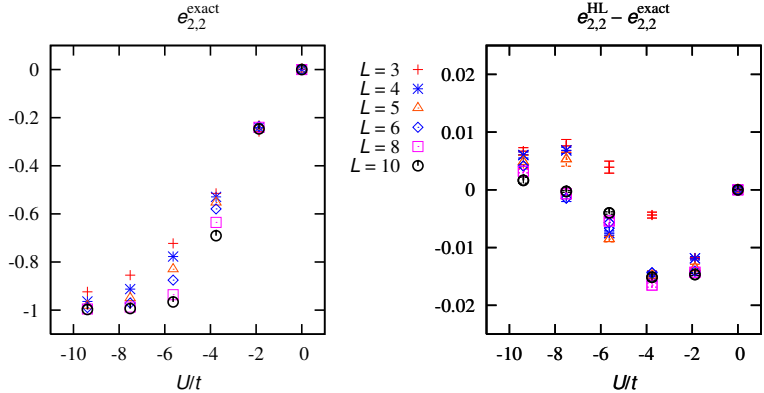}%
\caption{Comparison of four-body exact Lanczos results, $e_{2,2}%
^{\text{exact}}$, and heavy-light results, $e_{2,2}^{\text{HL}}$, for
$L=3,4,5,6,8,10$ lattices in two dimensions plotted versus coupling
$U/t=2mC.$}%
\label{twod_22}%
\end{center}
\end{figure}

\begin{figure}
[ptbptb]
\begin{center}
\includegraphics[
height=2.6048in,
width=4.4443in
]%
{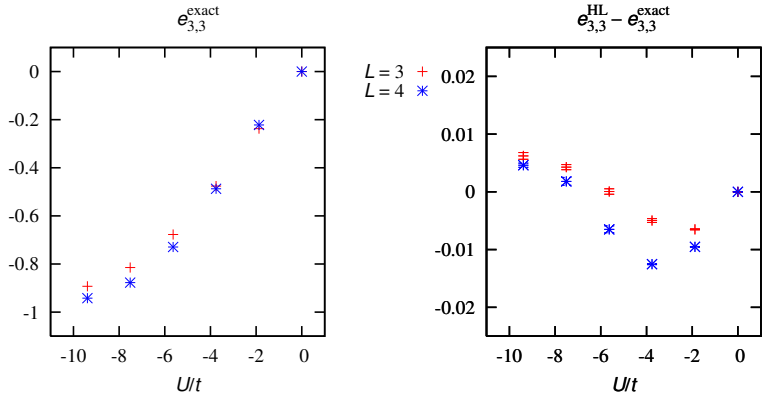}%
\caption{Comparison of six-body exact Lanczos results, $e_{3,3}^{\text{exact}%
}$, and heavy-light results, $e_{3,3}^{\text{HL}}$, for $L=3,4$ lattices in
two dimensions plotted versus coupling $U/t=2mC.$}%
\label{twod_33}%
\end{center}
\end{figure}

Fig.~\ref{twod_22} shows $e_{2,2}$ for the unpolarized four-body system in two
dimensions.%
Fig.~\ref{twod_33} shows $e_{3,3}$ for the unpolarized six-body system in two
dimensions.%
In both cases the error of the heavy-light result is bounded by $0.02$. \ Just
as in the one- and three-dimensional examples, the method appears to
accurately describe the crossover from fermions to condensed bosonic dimers.
\ In two dimensions the dimer wavefunction at infinite volume has an
exponential tail with characteristic length%
\begin{equation}
\frac{1}{\sqrt{mB_{2}}}=\frac{1}{1.80\pi}\exp\left(  -\frac{2\pi}{mC}\right)
=\frac{1}{1.80\pi}\exp\left(  -\frac{4\pi}{U/t}\right)  .
\end{equation}
The dimer increases its size dramatically for small negative $U/t$. \ But at
finite volume it eventually wraps around the periodic boundary at distance
$L$. $\ $This boundary effect changes the behavior of $e_{N,N}$. \ This can be
seen as a crossover to approximately linear dependence on $U/t$ for
$U/t\gtrsim-4$ in both the four-body and six-body results.

\section{Dimer-dimer scattering in three dimensions}

We have already presented results for $e_{2,2}$ in three dimensions. \ In this
section we reanalyze the data to extract low-energy scattering parameters for
dimer-dimer scattering. \ L\"{u}scher's formula
\cite{Luscher:1986pf,Beane:2003da,Seki:2005ns,Borasoy:2006qn} relates the
energy levels for any two-body system in a finite periodic cube to the S-wave
phase shift. \ In the appendix we discuss how to use L\"{u}scher's formula in
the two-body system with one up spin and one down spin to measure the
fermion-fermion scattering length, $a_{\text{scatt}}$. \ At strong coupling we
can also use L\"{u}scher's formula in the four-body system to measure the
S-wave dimer-dimer phase shift. \ When $L$ is much bigger than
$a_{\text{scatt}}$ we can interpret the energy difference%
\begin{equation}
\Delta E_{2,2}^{0}=E_{2,2}^{0}-2E_{1,1}^{0}%
\end{equation}
as the energy of the two-dimer system relative to threshold. \ We then
determine the dimer-dimer phase shift using%
\begin{equation}
p_{\text{D}}\cot\delta_{\text{DD}}=\frac{1}{\pi L}S\left(  \eta\right)
,\qquad\eta=\left(  \frac{Lp_{\text{D}}}{2\pi}\right)  ^{2},
\end{equation}
where $p_{\text{D}}$ is the dimer momentum, $S\left(  \eta\right)  $ is the
three-dimensional zeta function defined in the appendix, and $\delta
_{\text{DD}}$ is the dimer-dimer S-wave phase shift. \ It is convenient to
measure everything in units of the fermion-fermion scattering length
$a_{\text{scatt}}$.

\begin{figure}
[ptb]
\begin{center}
\includegraphics[
height=2.8236in,
width=2.9628in
]%
{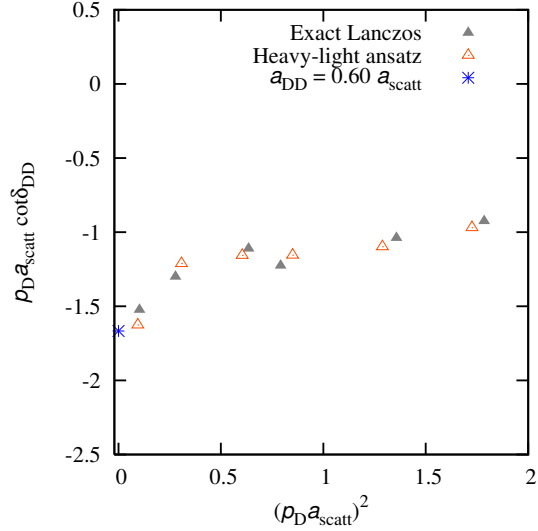}%
\caption{Results for $p_{\text{D}}\cot\delta_{\text{DD}}$ as a function of the
dimer momentum squared. \ Both quantities are measured in units of
$a_{\text{scatt}}$. \ For comparison we also show the limit of $p_{\text{D}%
}\cot\delta_{\text{DD}}$ at zero momentum determined by the result
$a_{\text{DD}}\approx0.60a_{\text{scatt}}$.}%
\label{kcot}%
\end{center}
\end{figure}

Results for $p_{\text{D}}\cot\delta_{\text{DD}}$ are shown in Fig.~\ref{kcot}
as a function of dimer momentum squared. \ For comparison we also indicate the
result $a_{\text{DD}}\approx0.60a_{\text{scatt}}$
\cite{petrov:2004PRL,Rupak:2006jj} which determines $p_{\text{D}}\cot
\delta_{\text{DD}}$ at zero momentum using the effective range expansion,%
\begin{equation}
p_{\text{D}}\cot\delta_{\text{DD}}\approx-\frac{1}{a_{\text{DD}}}+\frac{1}%
{2}r_{\text{DD}}p_{\text{D}}^{2}+\cdots.
\end{equation}%
We see in Fig.~\ref{kcot} that the heavy-light and exact Lanczos results agree
at the level of a few percent level. \ Also both agree with the dimer-dimer
scattering length result $a_{\text{DD}}\approx0.60a_{\text{scatt}}$. \ From
this plot we can also estimate the dimer-dimer effective range,%
\begin{equation}
r_{\text{DD}}\approx2.6a_{\text{scatt}}\text{.}%
\end{equation}

\section{BCS-BEC\ crossover in three dimensions}

\subsection{Many-body results at unitarity}

\begin{figure}
[ptb]
\begin{center}
\includegraphics[
height=3.0234in,
width=4.1131in
]%
{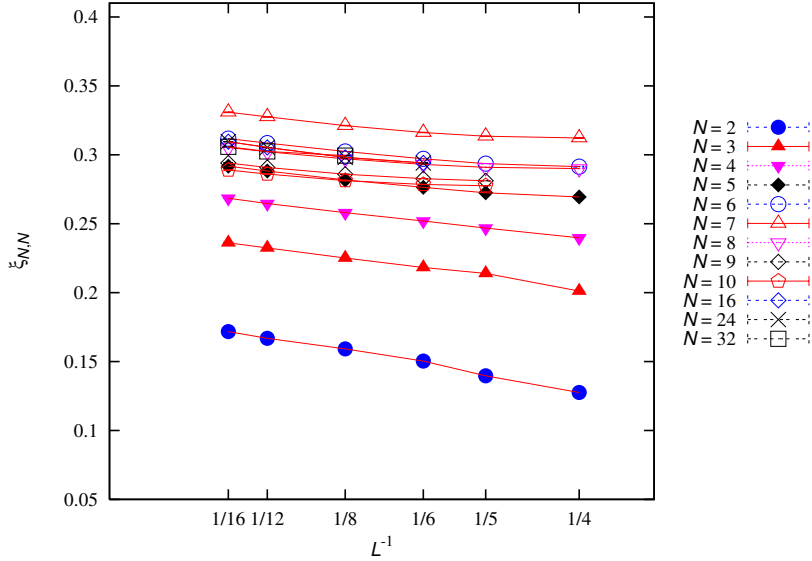}%
\caption{Heavy-light results for $\xi_{N,N}$ for $N$ from $2$ to $32$. \ The
results are plotted versus $L^{-1}$, where $L$ ranges from $4$ to $16$.}%
\label{ldependence_markov}%
\end{center}
\end{figure}

We present symmetric heavy-light results in the lowest filling approximation
at unitarity for%
\begin{equation}
\xi_{N,N}=\frac{E_{N,N}^{0}}{E_{N,N}^{0,\text{free}}}%
\end{equation}
for a wide range of values for $N$ and $L$.\ \ Fig.~\ref{ldependence_markov}
shows results for $\xi_{N,N}$ for the four-body system, $N=2$, up to the
sixty-four-body system, $N=32$. \ The results are plotted versus $L^{-1}$,
where $L$ ranges from $4$ to $16$. \ The dependence on $L$ is relatively mild.
\ The maximum for $\xi_{N,N}$ for $N=7$ appears to be caused by the closed
shell at $N=7$ for the free Fermi system in a periodic cube. \ In the
continuum and thermodynamic limits $L,N\rightarrow\infty$ we find
\begin{equation}
\xi=\lim_{L,N\rightarrow\infty}\xi_{N,N}=0.31(1).
\end{equation}%
The heavy-light results can be compared with published Euclidean lattice Monte
Carlo results at small volumes shown in Table 1 \cite{Lee:2005fk}.%
\[%
\genfrac{}{}{0pt}{0}{\text{Table 1: \ }\xi_{N,N}\text{ using Euclidean lattice
Monte Carlo}}{%
\begin{tabular}
[c]{|l|l|l|l|l|}\hline
$N=3$ & $N=5$ & $N=7$ & $N=9$ & $N=11$\\\hline
$0.19(2)$ & $0.24(2)$ & $0.28(2)$ & $0.23(2)$ & $0.25(2)$\\\hline
\end{tabular}
\ }%
\]
These results correspond with an average of data from $L=4,5,6$ for each
$N=3,5,7,9$ and $L=5,6$ for $N=11$. \ For the same $N$ and $L$, the
heavy-light results in Fig.~\ref{ldependence_markov} agree reasonably well
with the Euclidean Monte Carlo data, with a relative difference of at most
$20\%$.

The heavy-light data for larger values of $L$ probe much further towards the
continuum limit. \ These results suggest that lattice cutoff effects cannot
explain the discrepancy between lattice Monte Carlo results and continuum
fixed-node Monte Carlo results for the same values of $N$. \ Fixed-node
Green's function Monte Carlo calculations have found $\xi_{N,N}$ to be
$0.44(1)$ \cite{Carlson:2003z,Chang:2004PRA} and $0.42(1)$
\cite{Astrakharchik:2004} for comparable values of $N$. \ Further studies are
needed to determine if different nodal surfaces can produce lower ground state
energies in fixed-node calculations. \ The symmetric heavy-light ansatz may be
useful in probing this question more deeply. \ In addition to the
aforementioned lattice Monte Carlo and fixed-node simulations there are a
number of other theoretical calculations
\cite{Engelbrecht:1997,Baker:1999dg,Heiselberg:1999,Perali:2004,Papenbrock:2005,Bulgac:2005a,Schafer:2005kg,JChen:2006,Nishida:2006a,Nishida:2006b,Veillette:2006,Arnold:2007,Juillet:2007a,Abe:2007ff,Bulgac:2008b}
and experimental measurements of $\xi$
\cite{O'Hara:2002,Bourdel:2003,Gehm:2003,Bartenstein:2004,Kinast:2005,Stewart:2006}
which span the range from about $0.2$ to $0.6.$

\subsection{Results for general scattering length}

We consider $\xi_{N,N}$ as a function of $k_{F}a_{\text{scatt}}$. \ In the
limit of strong attractive coupling and the thermodynamic limit $N\rightarrow
\infty$ at fixed density, we get%
\begin{equation}
\xi=\lim_{N\rightarrow\infty}\frac{E_{N,N}^{0}}{E_{N,N}^{0,\text{free}}}%
=\lim_{N\rightarrow\infty}\frac{NE_{1,1}^{0}+\frac{N(N-1)}{2}\frac{4\pi
a_{\text{DD}}}{2mL^{3}}}{2\times\frac{3}{5}NE_{F}}+O(k_{F}^{2}a_{\text{scatt}%
}^{2}), \label{strongcoupling}%
\end{equation}
where $E_{1,1}^{0}$ is the energy for one dimer and $a_{\text{DD}}$ is the
dimer-dimer scattering length. \ The first term takes into account the binding
energy of the dimer while the second gives the contribution due to dimer-dimer
interactions. \ Although we have written the expansion in powers of
$k_{F}a_{\text{scatt}}$, a more accurate estimate of the appropriate expansion
parameter is $a_{\text{scatt}}$ divided by the average spacing between
particles $d\approx(6\pi^{2})^{1/3}k_{F}^{-1}$. \ We expect the $O(k_{F}%
^{2}a_{\text{scatt}}^{2})$ error to be small for $k_{F}a_{\text{scatt}%
}\lesssim2$.

For a periodic cube we have%
\begin{equation}
E_{F}=\frac{1}{2m}\left(  6\pi^{2}\frac{N}{L^{3}}\right)  ^{2/3}=\frac{\left(
6\pi^{2}N\right)  ^{2/3}}{2mL^{2}}%
\end{equation}
and%
\begin{equation}
k_{F}=\frac{\left(  6\pi^{2}N\right)  ^{1/3}}{L}.
\end{equation}
In the continuum limit the energy for one dimer is%
\begin{equation}
E_{1,1}^{0}=-\frac{1}{ma_{\text{scatt}}^{2}}, \label{dimer_continuum}%
\end{equation}
and the dimer-dimer scattering length is approximately $a_{\text{DD}}%
\approx0.60a_{\text{scatt}}.$ \ Putting these together we get%
\begin{equation}
\xi=\lim_{N\rightarrow\infty}\frac{E_{N,N}^{0}}{E_{N,N}^{0,\text{free}}%
}=-\frac{5}{3k_{F}^{2}a_{\text{scatt}}^{2}}+0.60\times\frac{5}{18\pi}%
k_{F}a_{\text{scatt}}+O(k_{F}^{2}a_{\text{scatt}}^{2}).
\end{equation}
We refer to the first two terms in this expansion as leading order (LO) and
next-to-leading order (NLO) in the strong-coupling expansion.

\begin{figure}
[ptb]
\begin{center}
\includegraphics[
height=3.026in,
width=3.3875in
]%
{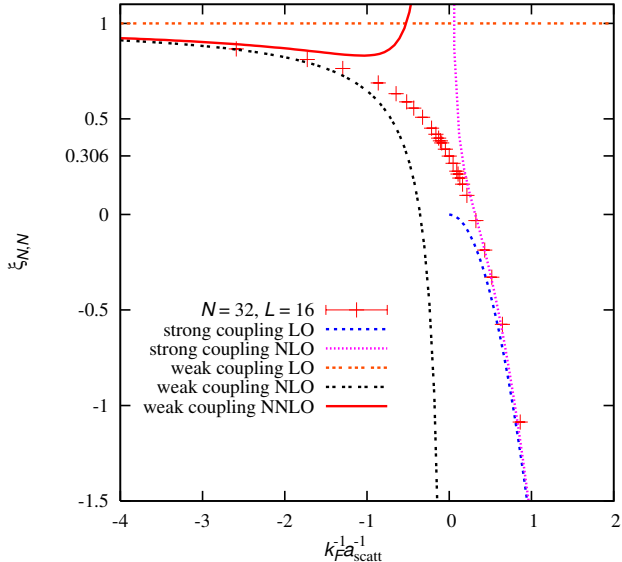}%
\caption{Plot of $\xi_{N,N}$ for $N=32$ and $L=16$ as a function of
$k_{F}^{-1}a_{\text{scatt}}^{-1}$. \ For comparison we show the analytic
strong-coupling and weak-coupling results.}%
\label{kfa}%
\end{center}
\end{figure}

In the weak-coupling limit we have \cite{Huang:1957A,Lee:1957A}%
\begin{equation}
\xi_{N,N}=\frac{E_{N,N}^{0}}{E_{N,N}^{0,\text{free}}}=1+\frac{10}{9\pi}%
k_{F}a_{\text{scatt}}+\frac{4(11-2\ln2)}{21\pi^{2}}k_{F}^{2}a_{\text{scatt}%
}^{2}+O(k_{F}^{3}a_{\text{scatt}}^{3}).
\end{equation}
We refer to the first three terms in this expansion as leading order (LO),
next-to-leading order (NLO), and next-to-next-to-leading order (NNLO) in the
weak-coupling expansion. \ In Fig.~\ref{kfa} we show $\xi_{N,N}$ as a function
of $k_{F}^{-1}a_{\text{scatt}}^{-1}$ for $N=32$ and $L=16$ using the symmetric
heavy-light ansatz in the lowest filling approximation. \ For comparison we
show the analytic strong-coupling and weak-coupling results.

We see that the heavy-light results are very close to the strong-coupling
results for $k_{F}^{-1}a_{\text{scatt}}^{-1}\gtrsim0.3$. \ The lowest filling
approximation is also not bad in the weak-coupling limit for $k_{F}%
^{-1}a_{\text{scatt}}^{-1}\lesssim-1$.%
Near the unitarity point we can expand%
\begin{equation}
\xi_{N,N}=\frac{E_{N,N}^{0}}{E_{N,N}^{0,\text{free}}}=\xi-\xi_{1}k_{F}%
^{-1}a_{\text{scatt}}^{-1}-\xi_{2}k_{F}^{-2}a_{\text{scatt}}^{-2}+\cdots.
\end{equation}
For $N=32$ and $L=16$ we find by least squares fitting that%
\begin{equation}
\xi=0.306(1),
\end{equation}%
\begin{equation}
\xi_{1}=0.805(2),
\end{equation}%
\begin{equation}
\xi_{2}=0.63(3).
\end{equation}
The heavy-light result for $\xi_{1}$ is within $20\%$ of a Euclidean lattice
Monte Carlo calculation \cite{Lee:2006hr} which found $\xi_{1}=1.0(1)$. \ In
contrast with $\xi$, there is general agreement in the recent literature on
the value of $\xi_{1}$ calculated using different methods
\cite{Chang:2004PRA,Astrakharchik:2004,Chen:2006A,Abe:2007ff}.

\section{Discussion}

\subsection{Lowest filling approximation and beyond}

In this analysis we have tested the symmetric heavy-light ansatz in the lowest
filling approximation on numerous few- and many-body systems for the 1D, 2D,
and 3D attractive Hubbard models. \ In each case the method appears to be
accurate. \ It is perhaps not surprising that the lowest filling approximation
is reliable at strong attractive coupling. \ In this limit the fermions form
tightly-bound dimers which are weakly interacting and condense in the ground
state. \ In the ground state the relative momentum between dimer pairs is
relatively low, and the up-spin coordinates $\mathbf{R}$ can be viewed as
surrogates for the locations of dimer pairs. \ The $N$-body fixed-point
constraint sets $\left\vert F_{\mathbf{R}}\right\vert ^{2}$ so that up spins
and down spins are arranged in the same manner relative to neighboring
particles. \ At very strong coupling $\left\vert F_{\mathbf{R}}\right\vert
^{2}$ is essentially independent of $\mathbf{R}$, indicating that the dimers
are non-interacting bosons each in the zero momentum state. \ As the
attractive coupling becomes weaker, the size of the dimers increases and the
effect of Fermi repulsion becomes more important.

For the systems we have considered here the lowest filling approximation works
well at weak attractive coupling. \ However in future studies one may consider
systematic improvements to the lowest filling approximation that include
excited orbitals. \ This improvement would probably be needed to study the
repulsive Hubbard model where there is no gap between the lowest $N$ orbitals
and higher orbitals. \ In the lowest filling approximation the coefficients
$\left\vert F_{\mathbf{R}}\right\vert ^{2}$ are set by imposing $Z_{2}$
symmetry on the same-spin correlations,%
\begin{equation}
\left\langle \Phi_{N,N}\right\vert :\rho_{\uparrow}(\vec{r}_{1})\times
\cdots\times\rho_{\uparrow}(\vec{r}_{N}):\left\vert \Phi_{N,N}\right\rangle
=\left\langle \Phi_{N,N}\right\vert :\rho_{\downarrow}(\vec{r}_{1}%
)\times\cdots\times\rho_{\downarrow}(\vec{r}_{N}):\left\vert \Phi
_{N,N}\right\rangle .
\end{equation}
If excited orbitals are included, a larger set of coefficients $\left\vert
F_{\mathbf{R,}\left\{  j_{n}\right\}  }\right\vert ^{2}$ must be determined.
\ These can set by imposing $Z_{2}$ symmetry on mixed-spin correlations such
as%
\begin{align}
\left\langle \Phi_{N,N}\right\vert  &  :\rho_{\uparrow}(\vec{r}_{1})\times
\rho_{\uparrow}(\vec{r}_{2})\times\rho_{\downarrow}(\vec{r}_{3}):\left\vert
\Phi_{N,N}\right\rangle \nonumber\\
&  =\left\langle \Phi_{N,N}\right\vert :\rho_{\downarrow}(\vec{r}_{1}%
)\times\rho_{\downarrow}(\vec{r}_{2})\times\rho_{\uparrow}(\vec{r}%
_{3}):\left\vert \Phi_{N,N}\right\rangle .
\end{align}
It is not clear that an algorithm can be designed to solve these constraints a
priori, as we did for the lowest filling approximation. \ Therefore one
approach would be to determine some parameterization for $\left\vert
F_{\mathbf{R,}\left\{  j_{n}\right\}  }\right\vert ^{2}$ with unknown
coefficients setting the relative weight of each orbital set $\left\{
j_{n}\right\}  $ for fixed $\mathbf{R}$. \ These coefficients could then be
determined a posteriori by least squares fitting to the mixed-spin correlation constraints.

\subsection{$\mathbf{R}$-commuting operators and phases}

We define an $\mathbf{R}$-commuting operator as any operator which commutes
with the up-spin density $\rho_{\uparrow}(\vec{r})$ for all $\vec{r}$. \ Some
examples of $\mathbf{R}$-commuting operators include $H_{\text{HL}}$ and the
up-spin and down-spin density operators. \ For any $\mathbf{R}$-commuting
operator $O$ the ground state expectation value is%
\begin{equation}
\left\langle \Psi_{N,N}^{0}\right\vert O\left\vert \Psi_{N,N}^{0}\right\rangle
=\sum_{\mathbf{R}}\sum_{\left\{  j_{n}\right\}  ,\left\{  j_{n}^{\prime
}\right\}  }F_{\mathbf{R,}\left\{  j_{n}\right\}  }^{\ast}F_{\mathbf{R,}%
\left\{  j_{n}^{\prime}\right\}  }O_{\left\{  j_{n}\right\}  ,\left\{
j_{n}^{\prime}\right\}  }^{\mathbf{R}},
\end{equation}
where%
\begin{equation}
O_{\left\{  j_{n}\right\}  ,\left\{  j_{n}^{\prime}\right\}  }^{\mathbf{R}%
}=\left[  \left\langle \mathbf{R}\right\vert \otimes\bigwedge\limits_{n}%
\left\langle f_{j_{n}}(\mathbf{R})\right\vert \right]  O\left[  \left\vert
\mathbf{R}\right\rangle \otimes\bigwedge\limits_{n}\left\vert f_{j_{n}%
^{\prime}}(\mathbf{R})\right\rangle \right]  .
\end{equation}
Suppose $O$ is an $\mathbf{R}$-commuting $M$-body operator with $M\ll N$.
\ Due to the orthogonality of orbitals, the orbital sets $\left\{
j_{n}\right\}  $ and $\left\{  j_{n}^{\prime}\right\}  $ must be the same for
all orbitals left untouched by $O$. \ Because of this constraint the diagonal
elements $O_{\left\{  j_{n}\right\}  ,\left\{  j_{n}\right\}  }^{\mathbf{R}}$
are enhanced by powers of $N$ relative to the off-diagonal matrix elements
$O_{\left\{  j_{n}\right\}  ,\left\{  j_{n}^{\prime}\right\}  }^{\mathbf{R}}$.
\ For the special case $O=H_{\text{HL}}$ all the off-diagonal matrix elements
are in fact zero. \ If the weight of $\left\vert F_{\mathbf{R,}\left\{
j_{n}\right\}  }\right\vert ^{2}$ is dominated by orbital sets $\left\{
j_{n}\right\}  $ close to the lowest orbital filling $\left\{  1,\cdots
,N\right\}  $, we can approximate $O_{\left\{  j_{n}\right\}  ,\left\{
j_{n}\right\}  }^{\mathbf{R}}$ by the diagonal element at lowest orbital
filling,%
\begin{equation}
O_{\left\{  j_{n}\right\}  ,\left\{  j_{n}\right\}  }^{\mathbf{R}}\approx
O_{\left\{  1,\cdots,N\right\}  ,\left\{  1,\cdots,N\right\}  }^{\mathbf{R}}.
\end{equation}
This leaves us with%
\begin{equation}
\left\langle \Psi_{N,N}^{0}\right\vert O\left\vert \Psi_{N,N}^{0}\right\rangle
\approx\sum_{\mathbf{R}}O_{\left\{  1,\cdots,N\right\}  ,\left\{
1,\cdots,N\right\}  }^{\mathbf{R}}\sum_{\left\{  j_{n}\right\}  }\left\vert
F_{\mathbf{R,}\left\{  j_{n}\right\}  }\right\vert ^{2}\approx\sum
_{\mathbf{R}}O_{\left\{  1,\cdots,N\right\}  ,\left\{  1,\cdots,N\right\}
}^{\mathbf{R}}\left\vert F_{\mathbf{R}}\right\vert ^{2}.
\end{equation}

The expectation values of $\mathbf{R}$-commuting operators are simple since we
can keep only diagonal matrix elements, and the sign and phase of
$F_{\mathbf{R}}$ is irrelevant. \ The expectation values of other operators
are more challenging. \ One example that is not $\mathbf{R}$-commuting is the
difermion pair correlation,%
\begin{equation}
\psi^{2\dagger}(\vec{r})\psi^{2}(\vec{0}),
\end{equation}
where%
\begin{equation}
\psi^{2}(\vec{r})=a_{\uparrow}(\vec{r})a_{\downarrow}(\vec{r}).
\end{equation}
Here we need to compute matrix elements for orbitals from different up-spin
configurations as well as nontrivial geometric phases which may arise going
from one up-spin configuration $\mathbf{R}$ to another $\mathbf{R}^{\prime}$.

\subsection{Non-uniform systems}

The Markov chain algorithm for $N$-body fixed-point densities at lowest
filling works also for the non-uniform case with external potential $V(\vec
{r})$. \ Heavy-light calculations should be possible for systems such as
harmonic traps used in cold atomic experiments. \ In the case of
slowly-varying potentials these calculations could be compared with results
obtained using density functional theory
\cite{Papenbrock:2006hg,Engel:2006qu,Furnstahl:2007xm,Bulgac:2007a,Furnstahl:2008df}%
.

Let us define $H\left[  V\right]  $ with potential $V(\vec{r})$ as%
\begin{equation}
H\left[  V\right]  =H+\sum_{\vec{r}}V(\vec{r})\left[  \rho_{\uparrow}(\vec
{r})+\rho_{\downarrow}(\vec{r})\right]  ,
\end{equation}
where $H$ is the attractive Hubbard Hamiltonian defined in Eq.~(\ref{H}). \ If
$E_{N,N}^{0}[V]$ is the ground state energy of $H\left[  V\right]  $, the
straightforward generalization of the symmetric heavy-light ansatz gives%
\begin{equation}
\min_{U\left\vert \Phi_{N,N}\right\rangle =\left\vert \Phi_{N,N}\right\rangle
}\frac{\left\langle \Phi_{N,N}\right\vert H_{\text{HL}}\left[  V\right]
\left\vert \Phi_{N,N}\right\rangle }{\left\langle \Phi_{N,N}\right.
\left\vert \Phi_{N,N}\right\rangle }=E_{N,N}^{0}\left[  V\right]  ,
\label{ansatz2}%
\end{equation}
where%
\begin{equation}
H_{\text{HL}}\left[  V\right]  =H_{\text{HL}}+\sum_{\vec{r}}V(\vec{r})\left[
\rho_{\uparrow}(\vec{r})+\rho_{\downarrow}(\vec{r})\right]  .
\end{equation}

However we could also define a more general form for $H_{\text{HL}}$,%
\begin{equation}
H_{\text{HL}}\left[  V,V_{A}\right]  =H_{\text{HL}}+\sum_{\vec{r}}\left[
V(\vec{r})-V_{A}(\vec{r})\right]  \rho_{\uparrow}(\vec{r})+\sum_{\vec{r}%
}\left[  V(\vec{r})+V_{A}(\vec{r})\right]  \rho_{\downarrow}(\vec{r}),
\end{equation}
for arbitrary $V_{A}(\vec{r})$. \ The contribution of $V_{A}(\vec{r})$ cancels
from the expectation value in Eq.~(\ref{ansatz2}). \ For unpolarized systems
adding an overall constant to the auxiliary potential $V_{A}(\vec{r})$ has no
effect, and so we can set%
\begin{equation}
\sum_{\vec{r}}V_{A}(\vec{r})=0.
\end{equation}
The optimal $V_{A}(\vec{r})$ can be found by minimizing the Rayleigh-Ritz
ratio,%
\begin{equation}
\min_{U\left\vert \Phi_{N,N}\right\rangle =\left\vert \Phi_{N,N}\right\rangle
}\frac{\left\langle \Phi_{N,N}\right\vert H_{\text{HL}}\left[  V,V_{A}\right]
\left\vert \Phi_{N,N}\right\rangle }{\left\langle \Phi_{N,N}\right.
\left\vert \Phi_{N,N}\right\rangle },
\end{equation}
in the lowest filling approximation. \ Roughly speaking the $V_{A}(\vec{r})$
adjusts the single-particle down-spin density,%
\begin{equation}
\left\langle \Phi_{N,N}\right\vert :\rho_{\downarrow}(\vec{r}):\left\vert
\Phi_{N,N}\right\rangle ,
\end{equation}
while the $N$-body fixed-point constraint insures that the single-particle
densities are equal for both spins,%
\begin{equation}
\left\langle \Phi_{N,N}\right\vert :\rho_{\uparrow}(\vec{r}):\left\vert
\Phi_{N,N}\right\rangle =\left\langle \Phi_{N,N}\right\vert :\rho_{\downarrow
}(\vec{r}):\left\vert \Phi_{N,N}\right\rangle .
\end{equation}
Applications of this approach to non-uniform systems will be discussed in a
future publication.

\section{Summary}

We have presented a many-body approach called the symmetric heavy-light
ansatz. \ It is an approximate method for finding the ground state energy of
dilute unpolarized two-component fermions with attractive interactions.
\ Although the Hamiltonian has an exact $Z_{2}$ symmetry, the ansatz breaks
the symmetry by changing the ratio of the masses of the two components. \ We
considered the extreme limit where one component is infinitely heavy and the
many-body problem can be solved in terms of single-particle orbitals. \ The
original $Z_{2}$ symmetry was reintroduced by setting the $N$-body density
correlations for the two components equal for all $N$. \ A Markov chain
algorithm was designed to generate exactly this $N$-body density constraint.

To test the method we first compared the results of the symmetric heavy-light
ansatz with exact Lanczos results for the four-body system in three
dimensions. \ We then tested the method with exact results for four- and
six-body systems in both one and two dimensions. \ We then considered
dimer-dimer scattering and larger systems in three dimensions at unitarity and
arbitrary values for the scattering length. \ The results indicate that the
method is quite accurate and robust. \ We discussed some extensions beyond the
lowest filling approximation and applications to non-uniform systems.

\section{Acknowledgements}

The author thanks Gautam Rupak and Thomas Sch\"{a}fer for discussions and
comments. \ This work is supported in part by DOE grant DE-FG02-03ER41260.

\appendix

\section{Finite-volume scattering in three dimensions}

In this appendix we use L\"{u}scher's formula
\cite{Luscher:1986pf,Beane:2003da,Seki:2005ns,Borasoy:2006qn} to relate the
coefficient $C$ in the three-dimensional Hubbard model to the S-wave
scattering length. \ We consider one up-spin particle and one down-spin
particle in a periodic cube of length $L$. \ L\"{u}scher's formula relates the
two-particle energy levels in the center-of-mass frame to the S-wave phase
shift,%
\begin{equation}
p\cot\delta_{0}(p)=\frac{1}{\pi L}S\left(  \eta\right)  ,\qquad\eta=\left(
\frac{Lp}{2\pi}\right)  ^{2}, \label{lusch}%
\end{equation}
where $S(\eta)$ is the three-dimensional zeta function,%
\begin{equation}
S(\eta)=\lim_{\Lambda\rightarrow\infty}\left[  \sum_{\vec{n}}\frac
{\theta(\Lambda^{2}-\vec{n}^{2})}{\vec{n}^{2}-\eta}-4\pi\Lambda\right]  .
\end{equation}
For $\left\vert \eta\right\vert <1$ we can expand in powers of $\eta$,%
\begin{align}
S(\eta)  &  =-\frac{1}{\eta}+\lim_{\Lambda\rightarrow\infty}\left[  \sum
_{\vec{n}\neq\vec{0}}\frac{\theta(\Lambda^{2}-\vec{n}^{2})}{\vec{n}^{2}-\eta
}-4\pi\Lambda\right] \nonumber\\
&  =-\frac{1}{\eta}+S_{0}+S_{1}\eta^{1}+S_{2}\eta^{2}+S_{3}\eta^{3}\cdots,
\end{align}
where%
\begin{equation}
S_{0}=\lim_{\Lambda\rightarrow\infty}\left[  \sum_{\vec{n}\neq\vec{0}}%
\frac{\theta(\Lambda^{2}-\vec{n}^{2})}{\vec{n}^{2}}-4\pi\Lambda\right]  ,
\end{equation}%
\begin{equation}
S_{j}=\sum_{\vec{n}\neq\vec{0}}\frac{1}{\left(  \vec{n}^{2}\right)  ^{j+1}%
}\qquad j\geq1.
\end{equation}
The first few coefficients are
\begin{align}
S_{0}  &  =-8.913631,\quad S_{1}=16.532288,\quad S_{2}=8.401924,\quad
S_{3}=6.945808,\nonumber\\
S_{4}  &  =6.426119,\quad S_{5}=6.202149,\quad S_{6}=6.098184,\quad
S_{7}=6.048263.
\end{align}
For small momenta the effective range expansion gives%
\begin{equation}
p\cot\delta_{0}(p)\approx-\frac{1}{a_{\text{scatt}}}+\frac{1}{2}r_{0}%
p^{2}+\cdots\text{,} \label{effrange}%
\end{equation}
where $a_{\text{scatt}}$ is the scattering length and $r_{0}$ is the effective range.

\begin{figure}
[ptb]
\begin{center}
\includegraphics[
height=0.8562in,
width=3.5276in
]%
{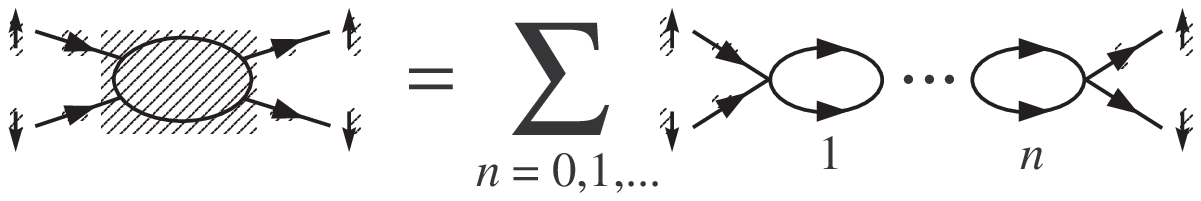}%
\caption{Sum of bubble diagrams contributing to two-particle scattering}%
\label{twotwo}%
\end{center}
\end{figure}

In terms of $\eta$, the energy of the two-body scattering state is%
\begin{equation}
E_{\text{pole}}=\frac{p^{2}}{m}=\frac{\eta}{m}\left(  \frac{2\pi}{L}\right)
^{2}. \label{Epole}%
\end{equation}
We compute the two-particle scattering pole at energy $E_{\text{pole}}$ by
summing the two-particle bubble diagrams shown in Fig.~\ref{twotwo}.%
For spin-dependent masses $m_{\uparrow}$ and $m_{\downarrow}$ the
single-particle dispersion relations are%
\begin{align}
\omega_{\uparrow}(\vec{p})  &  =\frac{1}{m_{\uparrow}}\sum_{l}\left(  1-\cos
p_{l}\right)  ,\\
\omega_{\downarrow}(\vec{p})  &  =\frac{1}{m_{\downarrow}}\sum_{l}\left(
1-\cos p_{l}\right)  .
\end{align}
After summing the bubble diagrams as a geometric series, the relation between
$C$ and $E_{\text{pole}}$ is%
\begin{equation}
-\frac{1}{C}=\lim_{L\rightarrow\infty}\frac{1}{L^{3}}\sum_{\vec{k}\text{
}\operatorname{integer}}\frac{1}{-E_{\text{pole}}+\omega_{\uparrow}(2\pi
\vec{k}/L)+\omega_{\downarrow}(2\pi\vec{k}/L)}. \label{pole}%
\end{equation}
When combined with Eq.~(\ref{lusch}), (\ref{effrange}), and (\ref{Epole}), we
can relate $C$ and the scattering length $a_{\text{scatt}}$. \ We note that
for both $H$ and $H_{\text{HL}}$ we have%
\begin{equation}
\omega_{\uparrow}(2\pi\vec{k}/L)+\omega_{\downarrow}(2\pi\vec{k}/L)=\frac
{2}{m}\sum_{l}\left[  1-\cos\left(  2\pi k_{l}/L\right)  \right]  .
\end{equation}
Therefore the coefficient $C$ is exactly the same in both cases. \ The
unitarity limit corresponds with the value $U/t=2mC=-7.914$.

\bibliographystyle{apsrev}
\bibliography{NuclearMatter}

\end{document}